    \documentclass[10pt,onecolumn,twoside]{IEEEtran} 

\usepackage{multirow}
\usepackage{amsfonts}
\usepackage{epsfig}
\usepackage{amsmath}
\usepackage{amssymb}
\usepackage[nolist]{acronym}
\usepackage[english]{babel}
\usepackage{cite}
\usepackage{color}
\usepackage{stfloats}
\usepackage{algorithm}
\usepackage{algorithmic}
\usepackage{subfigure}

\newcommand{\mat}{\pmb}

\begin{document}

\begin{acronym}
\acro{DASH}{dynamic adaptive streaming over HTTP}  
\acro{CDN}{content delivery network}  
\acro{BP}{belief-propagation}  
\acro{CDF}{cumulative density function}
\acro{PDF}{probability density function}  
\acro{QoE}{quality of experience}   
\acro{ILP}{integer linear programming}   
\acro{CRDSA}{contention resolution diversity slotted ALOHA}
\acro{SA}{slotted ALOHA}
\acro{BS}{base station}
\acro{RD}{rate-distortion}
\acro{MAC}{medium access control}
\acro{IC}{interference cancellation}
\acro{IRSA}{irregular repetition slotted ALOHA}
\acro{EEP}{equal error protection}
\acro{UEP}{unequal error protection}
\acro{SIC}{successive interference cancellation}
\acro{BN}{burst node}
\acro{SN}{slot node} 
\end{acronym}

\title{ Prioritized Random MAC Optimization  via Graph-based  Analysis}
 
\author{
Laura Toni,~\IEEEmembership{Member,~IEEE},
Pascal Frossard,~\IEEEmembership{Senior Member,~IEEE}
\begin{small}
\thanks{L. Toni, and P. Frossard are with \'Ecole Polytechnique F\'ed\'erale de Lausanne (EPFL), Signal Processing Laboratory - LTS4, CH-1015 Lausanne, Switzerland. Email: \texttt{\{laura.toni, pascal.frossard\}@epfl.ch.} 
}
\thanks{This work was partially funded by the Swiss National Science Foundation (SNSF) under the CHIST- ERA project CONCERT (A Context-Adaptive Content Ecosystem Under Uncertainty), project nr. FNS 20CH21 151569.}
\end{small}
}%
\maketitle
\thispagestyle{empty}

\begin{abstract}
 Motivated by the  analogy between successive interference cancellation and iterative belief-propagation on erasure channels, irregular repetition slotted ALOHA (IRSA) strategies  have received a lot of attention in the design of  medium access control protocols. The IRSA schemes have been mostly analyzed for theoretical scenarios for homogenous sources, where they are shown to substantially improve the system performance compared to classical slotted ALOHA protocols. In this work, we consider generic systems where sources in different importance classes compete for a common channel. We   propose  a new   \emph{prioritized}   IRSA  algorithm and derive the probability to correctly resolve collisions for data from each source class.    We then make use of our theoretical analysis to formulate a new optimization problem for selecting the transmission strategies of heterogenous sources. We optimize  both the replication probability per class and the source rate per class, in such a way that the overall system utility  is maximized. We then propose  a heuristic-based algorithm  for the selection of the transmission strategy, which is  built on intrinsic characteristics of the iterative decoding methods adopted for recovering from collisions.   Experimental results   validate the accuracy of the theoretical study and show the gain of well-chosen  prioritized transmission strategies for transmission of data from heterogenous classes over shared wireless channels.
\end{abstract}

\section{Introduction}
\begin{figure}[t]
\begin{center}
\includegraphics[width=0.6\linewidth,  draft=false]{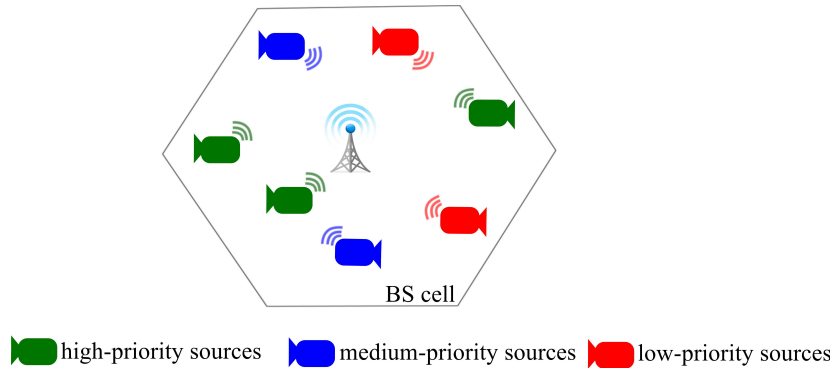}
\caption{Scenario with multiple sources (cameras) communicating to a central base station. Different levels of  priority could be given to the sources, with   high- medium- and low-priority   cameras, for example, which receive different shares of the network resources.  }   \label{fig:Scenario_general}
\end{center}
\end{figure}

In the era of   Internet of the Things, the number of   devices (sensors, machine terminal devices, portable devices, etc.) that are simultaneously connected to the network  is expected to grow very rapidly   in the near future~\cite{Zanella_zorzi:C13,Andrews:A14}. When a  massive number of devices share the same channel resources, there is an obvious need for   networks to be opportunistically designed with adaptive and distributed protocols.  In this context, random \ac{MAC} protocols have received a lot of attention since they do not require explicit coordination between wireless  network users.     At the same time, when different  classes of sources compete for a common channel, as illustrated in Fig. \ref{fig:Scenario_general}, a prioritized   allocation of available resources among sources is necessary in order to optimize the overall network utility.  The adoption of  \emph{prioritized random} MAC strategies  in future networks is thus desirable, creating the need for effective optimizations of multi-sources resource allocation strategies. 

The \ac{SA} protocol  has been widely considered as one effective random MAC strategy, where users randomly select the time slots where they transmit   information. If different users select the same time slot for transmission, a packet collision is experienced. While   collided packets were irremediably lost in early versions of  \ac{SA},  recent studies have shown that collisions  can be resolved by
 network diversity,  multiuser detection, network coding strategies \cite{wu2013,Cocco:J14,Cocco:Arxiv14}, or  by \ac{SIC} techniques \cite{Casini:J07} which substantially improves the system throughput. The key concept behind \ac{SIC} is that each user might send repetitions of the same message in different slots. If two messages from two different sources are sent in the same time slot (i.e., if a collision is experienced), the \ac{BS} might recover the messages through SIC if one of the collided messages has been decoded  previously. 
The throughput gain in applying SIC to SA schemes has been initially formalized in  \ac{CRDSA}, where     each user sends its own message within a MAC frame and  eventually a    replica   in a randomly selected slot \cite{Casini:J07}.  If a message is correctly decoded (with no collisions), it  can be used to  remove the potential interference contribution caused by the replicated  message. Using similar concepts,  authors proposed a novel random access protocol that exploits \ac{SIC} in a tree algorithm in \cite{YuGia:J07}.  A further analysis of such SIC-based random MAC protocol is proposed   in \cite{AndTur:C11}. 
   
An improvement of \ac{CRDSA}   has been proposed in \cite{Liva:J11}, where the author introduced   the  optimized transmission technique for \ac{IRSA} algorithm.  It consists in a random SA protocol  where the number of replicas that each user sends per frame (i.e., the replication rate) is not limited to two (or to any deterministic value) and it is rather  randomly selected according to a pre-determined transmission  probability distribution. A key connection is shown between the  SIC in IRSA and the  \ac{BP} decoder of   erasure codes on graphs \cite{Liva:J11}. This has opened the possibility of applying  theory of  rateless codes (and   codes on graph in general) to IRSA schemes  to optimize users' transmission strategies via proper selection of their   transmission  probability distribution \cite{Paolini_Arxiv:J14,SteVuk:L12,StePop:J13, PaoliniSLP14,MASSAP, Munari:J14,StefanovicMP14}. For example, the  IRSA has been   improved with coded slotted ALOHA schemes \cite{Paolini_Arxiv:J14}, as well as frameless slotted ALOHA protocols, in which the MAC frame size is not a priori selected \cite{SteVuk:L12, StePop:J13}.   However, to the best of our knowledge, none of the  recent works on random slotted ALOHA protocols with \ac{SIC} have considered unequal allocation of the channel resources among users. We exactly   aim at filling this gap and extend the IRSA framework to   prioritized random MAC protocols for heterogenous wireless sources.

 We  propose  a  \emph{prioritized     \ac{IRSA} algorithm}, in which  sources are  of different classes, which transmit information to a common BS with a random \ac{IRSA} strategy to access the channel. 
  Within a MAC frame, each source randomly  selects the replicas and the time slots to occupy,   independently from   other sources.   Each prioritized source class is identified by  an utility function and a transmission strategy. The utility function  is a non decreasing function of the received rate and it is such that sources in higher priority classes experience a larger utility score than users in lower priority classes for a given received rate. The transmission strategy  defined by the \ac{BS}    
  is characterized by the source rate and the transmission probability distribution (also defined as replication probability distribution) in IRSA, which is different in each class. 
 The transmission probability distribution drives the replication rate of sources in each class, hence the performance of the system. Our objective is to find the best transmission strategy that maximizes the expected utility over all classes.  Following the analogy between SIC and theory of  codes on graph   \cite{Liva:J11},  we analytically derive the probability   to correctly resolve collisions for data in each source class    along with the  expected utility per class.  Our theoretical analysis studies the performance of the iterative decoding algorithms for resolving collisions  in unequal transmission cases. It   resembles the AND-OR tree asymptotic analysis of LDPC codes over erasure channels.   We  exploit intrinsic characteristics of codes on graphs codes, i.e., the convergency of the iterative decoding method,   in order to derive decoding probability for random transmission probability distributions.   Because of the analogy between collision recovery schemes in IRSA and iterative message-passing algorithms on graph, we note that  the   UEP analysis of IRSA can be linked to UEP studies   for  irregular LDPC codes \cite{RahVelFek:J07,SejVukDouPie:J09,Sand:J10}.  However, there is a crucial  difference between the UEP MAC strategy considered in our work and    UEP rateless coding schemes. While  in the latter case the code designer controls the output nodes (check nodes) rather than the input nodes  (message nodes), this is exactly the opposite in  the  IRSA case, where the system designer controls the input nodes  (source nodes) rather than the output nodes (time slot nodes).   This  requires  a different analysis of the problem  compared to \cite{RahVelFek:J07}, for example.

Our new analysis  is then used to find the best unequal transmission strategy  among classes  in terms of both the replication probability and  source rate per class,  in such a way that the expected weighted utility   is maximized.    The underlying intuition is that more important classes should correctly receive  more messages than low priority classes. This is possible either by sending more messages from high-priority classes (i.e., by tuning the resource allocation strategy) or by sending messages with a transmission rate that guarantees  a lower failure  probability.    Results validate the accuracy of the theoretical study  and  show the gain of unequal transmission strategies for heterogenous  classes.   They also show  that  the proposed   method perform well when compared to optimal solutions computed by simulations.  

In summary, the main contributions of this work  are: 
\begin{itemize}
\item   a theoretical study of the  system performance in IRSA schemes with unequal transmission strategies, which leads to  the asymptotic message error probability per class as well as   global stability conditions; 
\item a new  optimization problem aimed at finding the best transmission strategy in prioritized IRSA strategies, in terms of both source rate and  replication probability  per class,  for a set of heterogenous  classes;  
\item   a solving method based on intrinsic characteristics of the iterative decoding method  adopted for recovering from collisions, along with proper  heuristics to derive random transmission strategies for different sources. 
\end{itemize} 
 
The reminder of this paper is organized as follows.  Section \ref{sec:system_model}    describes the scenario under consideration, together with key features of IRSA schemes. The theoretical analysis for prioritized IRSA strategies is derived in Section \ref{sec:error_prob}, where we also provide simulation results to validate the theory. The optimization problem aimed at finding the best transmission strategy for prioritized sources is formulated in Sec \ref{sec:SolvingMethod}. The solving method and simulation results are also provided. Finally,  we conclude in Section \ref{sec:conclusions}.

\section{Framework}
\label{sec:system_model}
In the following,  we first describe the system model considered in our work, then we detail key features of the SIC technique and   the IRSA framework.     
 
\subsection{System Model}
We consider  $M$ sources communicating to a common base station, that needs to decode the messages received from the sources. Sources are categorized in $K$ priority classes and we denote by $\mathcal{C}_k $ the set of active sources within the source class $k$. Let $L_k=|\mathcal{C}_k|$ be  the number of sources  in class $k$ with  $M = \sum_k L_k$.  Without loss of generality,   we assume that the source classes are sorted from the most important ($\mathcal{C}_1$) to the least one ($\mathcal{C}_K$).

\begin{figure}[t]
\begin{center}
\includegraphics[width=.9\linewidth,  draft=false]{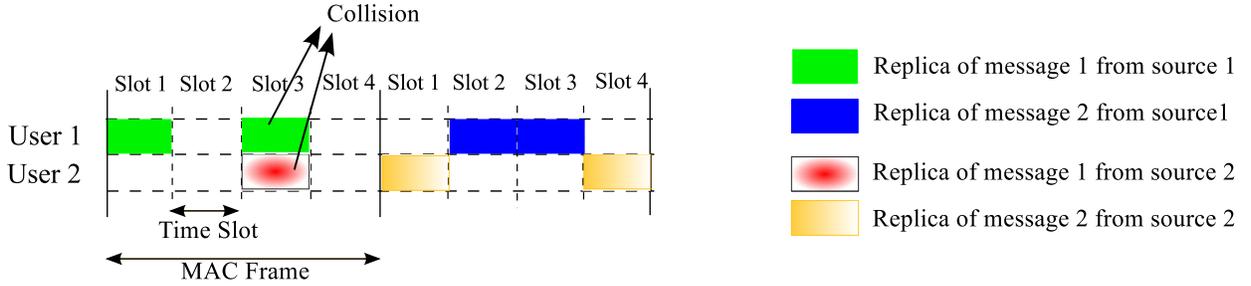}
\caption{Example of realization of IRSA strategies for two sources and four time slots per MAC frame. }\label{fig:MAC_Frame}
\end{center}
\end{figure}

We assume the time axis to be discretized in MAC frames of duration $T_{\text{MAC}}$ and we assume that at most one source packet is sent per source within a MAC frame. This means that at most $L_k$   messages can be sent per MAC frame from sources of class $k$.  
In our system, the sources access the channel according to the  \ac{IRSA} protocol~\cite{Liva:J11}. Each   MAC frame  of duration $T_{\text{MAC}}$ is composed of $N$ slots of duration $T_{\text{S}}=T_{\text{MAC}}/N$.  Each slot corresponds to a transmission interval, where one message or several interfering messages are sent. The traffic of the network is then computed as $G=M/N$. Within a MAC frame, each source  transmits $l$ replicas  of one source message, as depicted in Fig. \ref{fig:MAC_Frame}. 
Each replica   is transmitted within one time slot and replicas sent from the same source are allocated to different slots, which are uniformly selected at random among the $N$ total available slots. The replication rate $l$ is selected by the source at random   following a transmission probability distribution. We denote this distribution by $\{\Lambda_{l,k}\}_l$   for sources of class $k$, where $\Lambda_{l,k}$ is the probability that a source    from the   class  $k$   transmits $l$ replicas within the  MAC frame. 

\begin{table*}[t]
\centering
\scriptsize
\caption{Notation adopted in the performance analysis.}\label{tab:notation}
\begin{tabular}{l l}
\hline
Name & Description \\
\hline
 $T_S$ &  time slot over which a source can access the channel and send a replica message \\
$T_{MAC}$ & MAC frame duration\\ 
$N=T_{MAC}/T_S$ & transmission slots per MAC frame \\
$\mathcal{C}_k$ & set of sources of class $k$ \\
$L_k = |\mathcal{C}_k|$ & number of sources  assigned to class $k$\\
 $R_k$ & number of messages correctly received  in class $k$\\
  $U_k(R_k)$ & utility function for class $k$   \\
  \hline	
\end{tabular}
\end{table*}

The transmission processes are handled independently by all sources. This might lead to interference  on the wireless channel.  We assume that if a time slot is selected only by one source, the BS correctly receives the message.  When multiple sources select the same time slot for a replica transmission, a collision is experienced.  The messages interfere  and the information transmitted over the time slot cannot be  immediately recovered. However, the receiver implements successive interference cancellation (SIC)  to partially or fully resolve   collisions.    This is illustrated by an example in   Fig. \ref{fig:MAC_Frame}. In  the third time slot of the first MAC frame, the message sent by source $1$ ($m_1$) collides with the message of source $2$ ($m_2$). This means that the BS receives message $m_2$  interfered by $m_1$, making the messages undecodable.  However, thanks to  \ac{SIC} techniques, the   collision  might be resolved. In particular, the BS can recover $m_1$  from the time slot $1$, and once  $m_1$ is revealed, $m_2$  can be ``cleaned" from the interference with SIC algorithms. We assume that a perfect SIC is performed and the message is recovered with no errors \cite{Liva:J11}.  In a more general scenario, this interfering-cancelation procedure is iterated and may permit    the recovery of     the whole set of bursts transmitted within the same MAC frame.
We refer  readers to \cite{Liva:J11} for a  detailed description of SIC techniques applied to IRSA strategies.

 Finally, we   denote by  $R_k \in [0, \ldots, L_k]$ the number of  messages from class $k$ that are correctly received at the decoder. The reception of $R_k$ messages leads to an utility function $U_k(R_k)$, which  is a non-decreasing function of the rate.  Let denote by $\mat{R}=[R_1, R_2, \ldots, R_K]$   the vector of received rates  (or  received messages) for all classes. The overall utility function for the system is given by 
\begin{equation}\label{eq_exp_Dist} 
U(\mat{R}) = \sum_{k=1}^K  w_k U_k(R_k) 
\end{equation}  
where $w_k$ is a priority score that characterizes the importance of   class $k$, and it is such that $\sum_k w_k=1$. Usually important classes have large  weight $w_k$. 
The notation adopted in this work is summarized in Table \ref{tab:notation}.

\begin{figure}[t]
\begin{center}
\subfigure[Time slot representation]{
\includegraphics[width=0.5\linewidth,  draft=false]{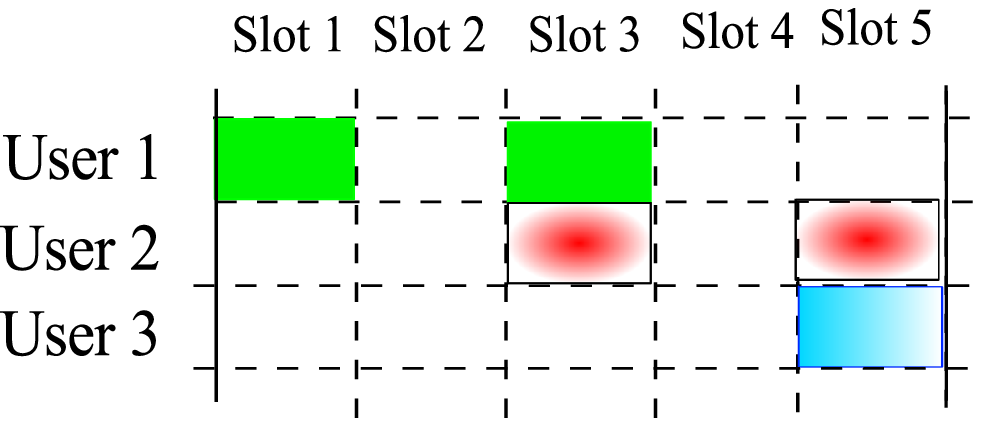}\label{fig:Graph_representation_A}}
\subfigure[Graph-based representation]{
\includegraphics[width=0.4\linewidth,  draft=false]{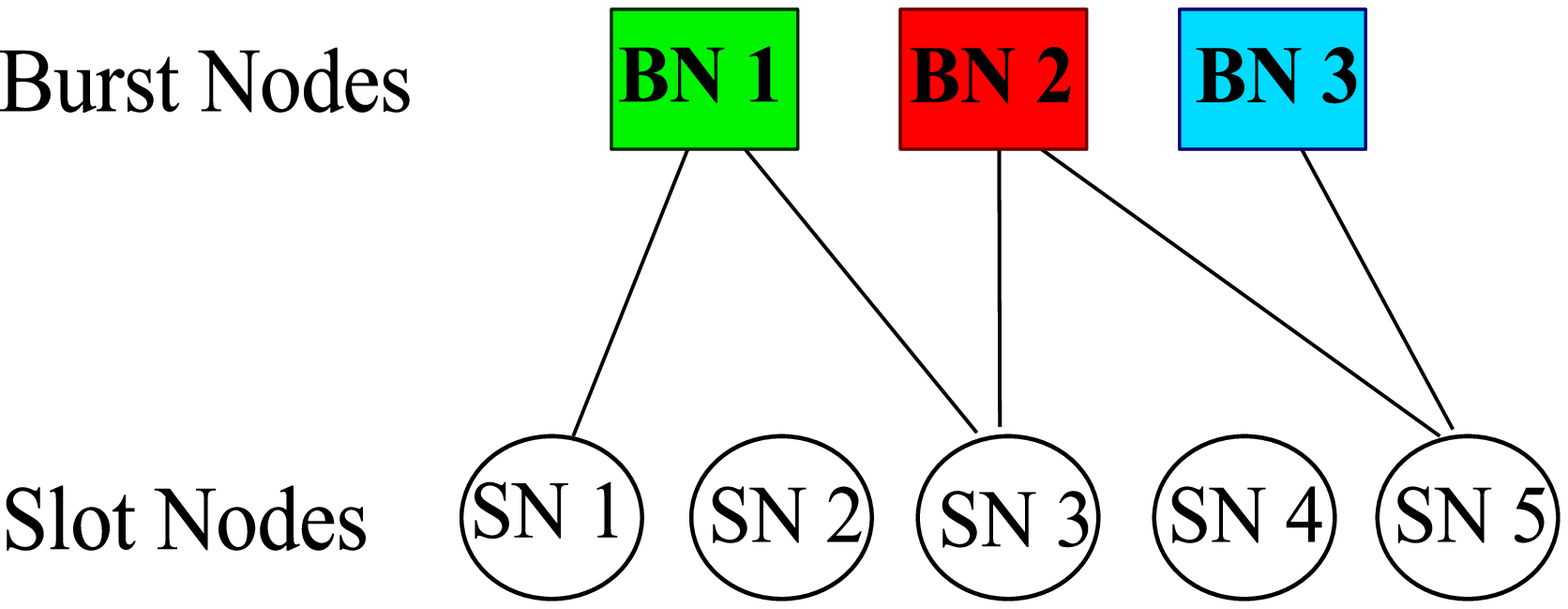}\label{fig:Graph_representation_B}}
\caption{Example of bipartite graph associated with the IRSA scheme, with $3$ sources of one message each.}\label{fig:Graph_representation}
\end{center}
\end{figure}
 
\subsection{Graph-based representation of IRSA} 
To study the performance of SIC strategies applied to IRSA protocols, a graph-based representation has been introduced in   \cite{Liva:J11}, which shows the analogy between random MAC protocols and codes on graph. Each MAC frame can be identified by a bipartite graph where sources are represented by \acp{BN}, that correspond to messages,  and transmission time slots are considered as \acp{SN}. The degree of a node defines the number of outgoing, respectively incoming edges.  The probability of having a degree-$l$ BN of class $k$ is given by $\Lambda_{l,k}$ and    the probability of having a degree-$l$ SN  is denoted by $\Omega_l$. 
The degree of a BN corresponds to the repetition rate adopted   by  the corresponding source  in the MAC frame. The degree of a SN corresponds to the number of interfering messages. Finally, the iterative message recovering procedure  is associated to a message-passing algorithm along the graph. An example of the bipartite graph representation is shown in Fig. \ref{fig:Graph_representation},  where  the MAC frame depicted in Fig. \ref{fig:Graph_representation_A}, characterized by $5$ time slots and $3$ sources with one message each, is identified by the bipartite graph in Fig. \ref{fig:Graph_representation_B}, with $5$ SNs and 3 BNs.  

\begin{figure}[t]
\begin{center}
\includegraphics[width=0.8\linewidth,  draft=false]{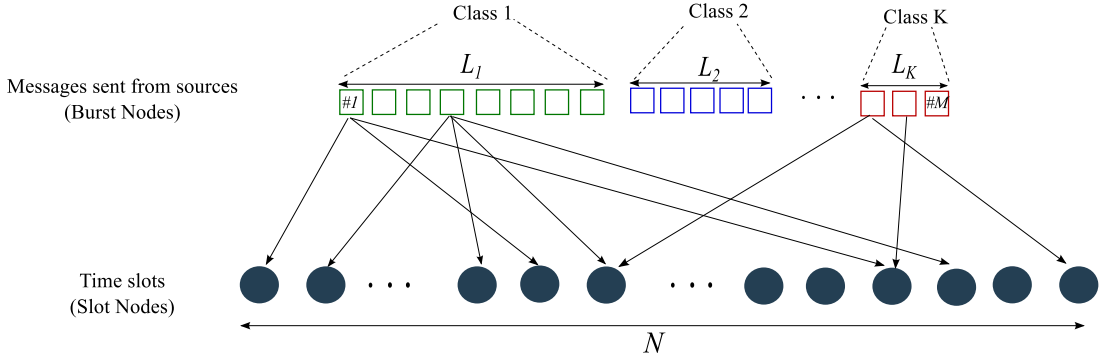}
\caption{Graph representation for the considered MAC protocol.   Sources of different classes   encode  the   information into messages that are sent in some of the $N$ time  slots selected at random.  }\label{fig:Camera_Bipartite}
\end{center}
\end{figure}

In  Fig. \ref{fig:Camera_Bipartite}, we extend the graph-based representation to the   scenario with heterogenous sources that is  considered in our work. Each source  in  $\mathcal{C}_k$   is identified by a \ac{BN} class $k$, which transmits message  replicas over the MAC frame independently from any other \acp{BN}. The frame status can then be represented by the bipartite graph $\mathcal{G}= (B,S,E)$ where $B$ is the set of $M$ burst nodes, $S$ is the set of $N$ \acp{SN}, and  $E$ is the set of edges. 
An edge  $(v,e)$ represents a transmission of BN $v$ in the time slot $e$.  In this case, we say that $v$ is a neighbor of $e$.    Recall that having a degree-$l$ BN of class $k$ corresponds to having  a source from class $k$ sending $l$  message replicas within  a MAC frame. Analogously, a degree-$l$ SN reflects a time slots in which $l$ messages have been transmitted.  
The   transmission strategy of the IRSA can then be identified by a  \emph{node-perspective} degree distribution  with the following polynomial representations 
\begin{align}
\Lambda_k(x) &= \sum_{l} \Lambda_{l,k} x^l ,  & \Omega(x)  = \sum_{l} \Omega_l x^l 
\end{align}
From the node-perspective  degree distributions, we can also derive the \emph{edge-perspective} degree distributions. We define  $\lambda_{l,k}$   the probability for an edge to be incident to a   degree-$l$ BN of class $k$  as follows 
\begin{align}
\lambda_{l,k} = \frac{ l \Lambda_{l,k}}{\sum_l l  \Lambda_{l,k}}\,.
\end{align}
Then, the edge-perspective degree distribution $\lambda(x)$ is given by 
\begin{align}\label{eq:edge_persp_lambda_2}
\lambda(x) = \sum_l \lambda_{l,k} x^{l-1} = \frac{\Lambda^{\prime}_k(x) }{\Lambda^{\prime}_k(1)}
\end{align}
where $\Lambda^{\prime}_k(x)= d\Lambda_k(x)  /dx$. 
Analogously, the probability of having an edge attached to a degree-$l$ SN is 
\begin{align}
\rho_{l} = \frac{ l \Omega_{l}}{\sum_l l  \Omega_{l}}\,.
\end{align}
Hence, the edge-perspective degree distribution $\rho(x)$ is given by
\begin{align}\label{eq:rho_exp_2}
  \rho(x) = \sum_l \rho_{l} x^{l-1} = \frac{\Omega^{\prime}(x) }{\Omega^{\prime}(1)}\,.
\end{align}
where $\Omega^{\prime}(x)= d\Omega(x)  /dx$.

 Different $(\Lambda_{k}(x), L_k)$  pairs lead to different frequency of accessing the channel and different transmission rates for the sources of  class $k$, creating an unequal allocation of the channels among different source classes.  In the following section, we show how the degree distribution is used to   derive a collision recovery probability per source class, which is the probability that a message sent from a source of a given class  is successively received.

\section{Collision Resolution Probability}\label{sec:error_prob}
We now evaluate the error probability for the prioritized IRSA schemes described above. 
The theoretical study   is evaluated under the assumption of   very large frame sizes ($N\rightarrow \infty$), hence the analysis presented next will refer to this asymptotic setting. 
The asymptotic assumption leads to theoretical analysis which is substantially simplified but yet accurate, as already proved in the literature \cite{Liva:J11, SteVuk:L12}. In the following, we will adopt  ``asymptotic setting" and ``large network assumption", interchangeably. 
 
In the following, first we  derive the SN degree distribution in the case of prioritized   transmission strategies. Then,  to derive the decoding error probability, we extend the asymptotic analysis in \cite{Liva:J11} to the   case of heterogenous sources.   Finally, we provide    global   conditions for the stability of the iterative decoding process.

\subsection{Node Degree Distributions}
In our scenario, the base station assigns to each   class $k$ a transmission strategy, which is defined by a transmission distribution $\Lambda_k(x)$. From $\Lambda_k(x)$, the edge-perspective distribution $\lambda_k(x)$ can be evaluated as in Eq.  \eqref{eq:edge_persp_lambda_2}. The degree distribution for the SNs, $\Omega(x)$ as well as $\rho(x)$, need to be computed. In \cite{Liva:J11}, the SN-degree distribution is derived for the case in which all BNs follow the same distribution $\Lambda(x)$. Here, we extend the analysis to   the case of different distributions  for different classes of sources, or equivalently  for different types of BNs. 
We denote by  $P_k(n_k)$ the probability that $n_k$ edges connect $n_k$ BNs of class $k$ to the same SN. This probability is given by 
\begin{align}
P_k(n_k)  &= {L_k \choose  n_k} p_k^{n_k} (1-p_k)^{L_k-n_k} \\
p_k &= \frac{\sum_j  j  \Lambda_{j,k}}{N}
\end{align}
where $p_k$  is the probability that  a BN  of class $k$ has an edge incident to the considered SN. Note that $p_k$ corresponds to the  probability that one  source from class $k$ transmits a replica  in the considered time slot. 
The degree distribution $\Omega_l$ for the SNs then becomes  
\begin{align}\label{eq:conv_prod}
\Omega_{l}  & = \sum_{n_1,\ldots,n_K: n_1+\ldots+n_K=l}    P_1(n_1) P_2(n_2) \ldots P_K(n_K) \\ \nonumber
  &=\sum_{n_1,\ldots,n_K: n_1+\ldots+n_K=l}    \prod_k {L_k \choose  n_k} \left(\frac{\sum_j j \Lambda_{j,k}}{N}\right)^{n_k} \left(1-\frac{\sum_j j \Lambda_{j,k}}{N}\right)^{L_k-n_k}, \ \ \ l=0,\ldots, M
\end{align}

We can simplify Eq.~\eqref{eq:conv_prod} as follows. We denote by $X_i$   the event of having the  source  $i$   transmitting in a time slot $t$. This event occurs with probability $p_i=(\sum_j j \Lambda_{j,k})/N$ if source  $i$ is from the class $k$. Since each source independently selects the time slots for transmission,   we have that  $X_1, X_2, \ldots, X_M$ are independent Bernoulli processes, each one with its own probability of success $p_i$.    Then $\Omega_{l}=P(S_M=l)$  can be modeled as a  Poisson binomial process \cite{wang1993}, where   $S_M=X_1+X_2 +\ldots + X_M$ is the sum of the considered events (i.e., $S_M$ sources transmitting in the time slot $t$).   This permits to express Eq. \eqref{eq:conv_prod}  as  
\begin{align}
\Omega_{l} = \sum_{A\in \mathcal{F}_l} \prod_{i\in A } (1-p_i)\prod_{i\in A^c} p_i 
\end{align}
where $\mathcal{F}_l$ is the set of all subsets of $l$ integers that can be selected in $M$, $A$ is one of these possible subsets, and $A^c$ is the complement of $A$, given by $\{1, 2, \ldots, M\} \setminus A$. 
From \cite{Fernandez:J10,hong2013}, the above expression becomes
\begin{align}
\Omega_{l} &= \frac{1}{M+1} \left\{  \sum_{j=1}^M {C}^{-jl} \left[ \prod_{m=1}^M \left( 
1 + (C^j-1)p_m
\right) \right]  \right\} 
\nonumber \\
&= \frac{1}{M+1} \left\{  \sum_{j=1}^M C^{-jl} \left[ \prod_{k=1}^K \left( 
1 + (C^j-1)p_k
\right)^{L_k} \right]  \right\} 
\end{align} 
where $C= \exp[2\pi i / (M+1)]$, with $i=\sqrt{-1}$. The last equality follows from the fact that sources from the same class have the same probability of success $p_k$. 
Finally,  for large $M$ and small $p_i$'s,  $\Omega_l$ can be approximated by a Poisson process  \cite{hodges1960}
\begin{align}
\Omega_{l} &\approx \frac{ \left( \sum\limits_{m=1}^M p_m \right)^l \exp\left(  - \sum\limits_{m=1}^M p_m \right) }{l!} =  \frac{ \left( \sum\limits_{k=1}^K  L_k p_k \right)^{l} \exp\left(  - \sum\limits_{k=1}^K  L_k p_k \right) }{l!} 
\end{align}
Note that we are considering large frame size networks ($N\rightarrow \infty$), which leads also to large $M$ values. Finally, each $p_i$ is inversely proportional to $N$ and it becomes small for large MAC frame size (i.e., large $N$). This then justifies the assumptions considered to derive the above approximation.

We can then derive the   node degree distribution for SNs as
\begin{align}\label{SN_degree_asympt}
\Omega(x) &= \sum_{l=1}^M \Omega_l x^l    
\approx  \exp(-\chi)\sum_{l=1}^M \frac{(\chi x)^l}{l!} 
\end{align}
where $\chi= \sum_{k=1}^K L_k p_k$. Under the assumption of large  networks, Eq. \eqref{SN_degree_asympt} can be further simplified  as  $\Omega(x) \approx \exp(-\chi(1-x))$.
Finally, from Eq. \eqref{eq:rho_exp_2}, the edge-perspective degree distribution for a SN   becomes  
\begin{align}\label{SN_edge_degree_asympt}
\rho(x) = \frac{\Omega^{\prime}(x)}{\Omega^{\prime}(1)}= \exp(-\lambda(1-x)) = \exp\left(- G \sum_k \frac{\Lambda_k^{\prime}(1) }{\sum_k L_k}  (1-x)\right)
\end{align}
where $G=\sum_k L_k/N$ is the traffic of the network.
 
\begin{figure}[t]
\begin{center}
\subfigure[Bipartite graph.]{
\includegraphics[width=0.38\linewidth,  draft=false]{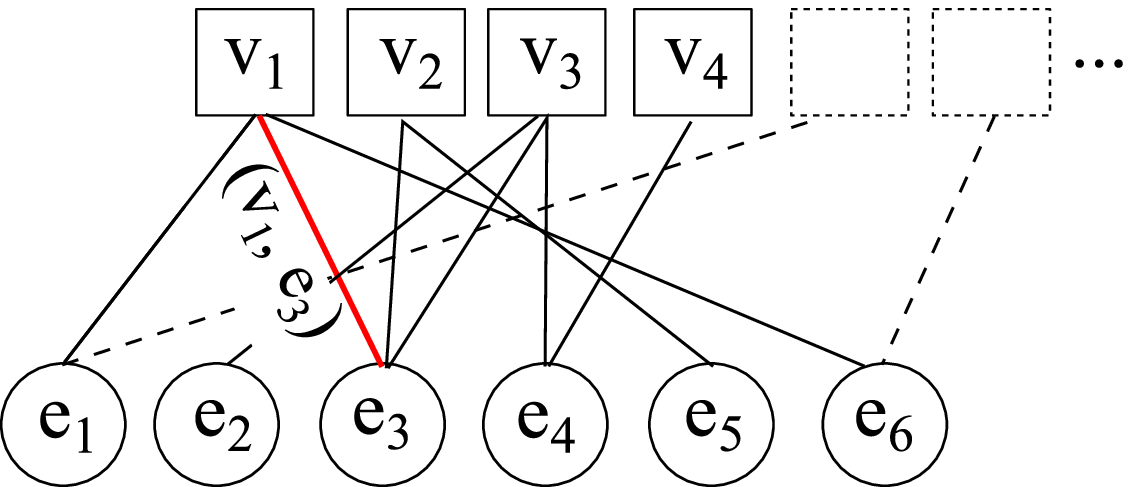}\label{fig:Tree_A}}
\hspace{2cm}
\subfigure[Tree  associated to $(v_1,e_3)$.]{
\includegraphics[width=0.23\linewidth, height=0.2\textheight , draft=false]{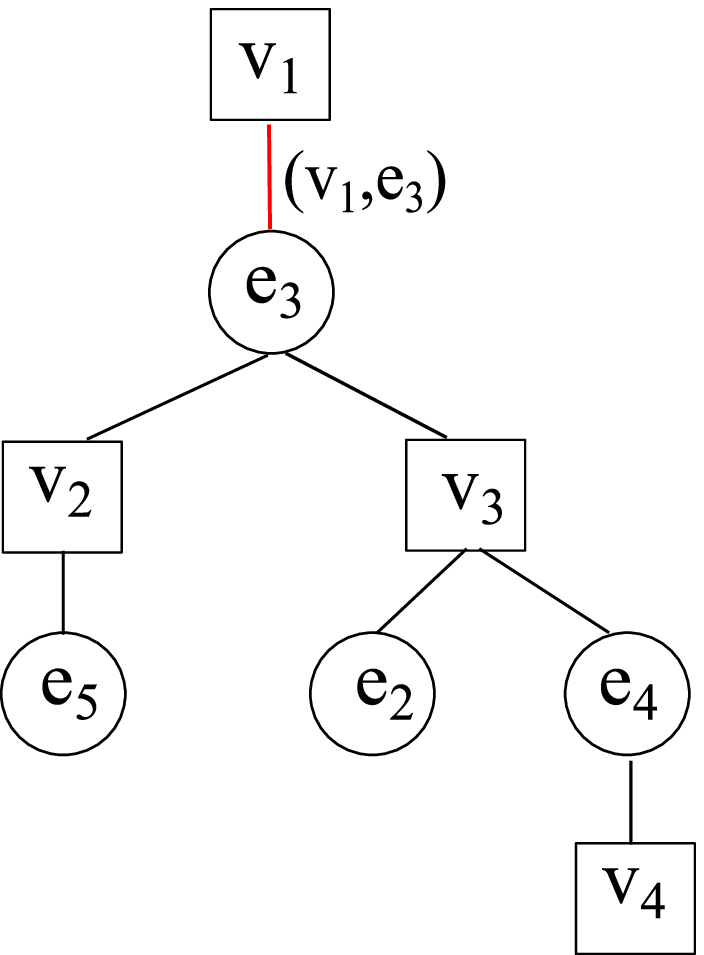}\label{fig:Tree_B}}
\caption{Example of the tree representation associated to $(v_1,e_3)$ in the bipartite graph, assuming that the BN $v_1$ is of class $k$ and it is the root of the tree $T_{i,k}$  of depth $2i = 4$.  }\label{fig:Tree}
\end{center}
\end{figure}

\subsection{Collision Recovery Probability}
\label{sec:Collision_Recovery_Probability}
We now study the collision recovery probability and  analyze the asymptotic behavior of the message-passing algorithm used in the SIC method. We extend the analysis in \cite{Liva:J11} and consider  the case of prioritized  transmission strategies.   The analysis considers the AND-OR tree asymptotic analysis, which is frequently used in evaluating  rateless codes performance \cite{Luby_andORtree:98,richardson:B08} and it  has been already introduced for IRSA strategies \cite{Liva:J11,SteVuk:L12}.  

First, we show the dependency among the nodes in the SIC algorithm and provide  an example in  Fig. \ref{fig:Tree_A}  that has two iterations in the decoder.   We see that  the slot node $e_5$ receives  a clean (i.e., not interfered) message  from $v_2$, which can then be decoded at the first step of the SIC procedure. We say that the edge $(v_2,e_5)$ is revealed at the first iteration step and it can be removed from the graph, leading to a   reduced graph.  The decoded message from $v_2$ is then passed along the edge $(v_2,e_3)$ to remove the interference at the third time slot of the MAC frame, namely $e_3$. This means that    the edge  $(v_2,e_3)$ is also   revealed and removed from the graph. If at the $i$th iteration step of the decoding  the message from the burst node $v_3$ has also been decoded (and so the message along $(v_3,e_3)$ has been passed), the message from $v_1$ can be decoded through the slot node $e_3$. 

Dependencies between nodes in the graph can be described by   a tree representation  \cite{Luby_andORtree:98}, which permits to   study  the iterative burst decoding process.  Considering   a burst node $v$, we are interested in the probability of decoding the message through a slot node $e$. The associated tree is the one  describing the  neighborhood of $e$.  This tree is rooted at  $v$, with $e$ as only branch going out from $v$, while the node $e$ has   branches to  the neighboring burst nodes  excluding $v$, as shown in Fig. \ref{fig:Tree} for the $(v_1,e_3)$ edge.

In more details, let denote by   $T_{i,k}$  the constructed tree of depth $2i$ with a BN of class $k$ (e.g., $v_1$) as a root.   Each node at depth $2i, 2i-2, \ldots, 2, 0$ are BNs (of  class  $k$ for the root and of any class for the other depths value) while the nodes at depth  $2i-1, 2i-3, \ldots, 3, 1$ are SNs. 
  Finally, the     nodes  at depth $2i, 2i-2, \ldots, 2, 0$ are denoted by OR-nodes, while the   nodes at depth $2i-1, 2i-3, \ldots, 3, 1$ are AND-nodes in the tree representation.  
In the tree representation, 
a BN at depth $2i$ (a SN at depth $2i-1$) is marked with $1$  if it is decoded (if it receives a clean message) at the $i$th iteration step, and marked with $0$ otherwise.    For the message sent from $v_1$   to be decoded at the $i$th decoding step through the slot node $e_3$, the slot node $e_3$ has to be marked with $1$.  This happens \emph{if all} the  other $(l-1)$ neighboring BNs of $e_3$ are decoded or marked with $1$  (\emph{AND}-operator), where $l$ is the degree of the slot node under consideration. We denote by $z_i$ the probability of the slot nodes at depth $2i-1$ to be marked  with $0$.  We consider now the probability that  any BN of degree $l$ at depth $i$ in the tree is marked with $1$. This happens  \emph{if at least one}    of the remaining $(l-1)$ neighboring SNs is marked with $1$, i.e., if it  has a non-interfered message (\emph{OR}-operator). 
We denote by $y_{i,k}$ the probability that the message along the considered edge is not decoded.  
%
%
In our IRSA scheme,  the messages transmitted by the sources are not known a priori and they need to be decoded. This means that none of the BNs is  known a priori  before the decoding process starts, i.e., $y_{0,k}=1, \forall k$ and $z_0=1$ and  all leafs are initially marked with $0$.

We now evaluate  the probability of having a BN of class $k$   unknown  after the $i$th iteration of the decoding process. This is given by 
\begin{align}
y_{i,k}  &= P\{ \text{all AND nodes at depth $2i-1$ are marked with 0}\} =Ê\sum_{l=1}^{N-1}  z_{i-1}^{l-1} \lambda_{l,k} =\lambda_{k}(z_{i-1}) \label{eq:lambda}  
\end{align}
where $z_{i-1}$ is the probability of having a AND-node at depth $2i-1$ that is marked with $0$ and $\lambda_{l,k}$ is the probability for an edge to be incident to a degree-$l$ BN of class $k$. Each child of a AND-node is a   OR-node of class $k$ with probability $q_k$, given by 
$$
q_k =  \frac{L_k \Lambda_{k}^{\prime}(1)}{\sum_{k=1}^K L_k \Lambda_{k}^{\prime}(1)}\, 
$$
 which is the ratio between the average number of edges going out from all BNs of class $k$, namely   $L_k \Lambda_{k}^{\prime}(1)$,  and the average number of total edges in the graph, namely $\sum_{k=1}^K L_k \Lambda_{k}^{\prime}(1)$.
It follows that  
\begin{subequations}
\begin{align}
z_{i-1} &= 1-    P\{ \text{all OR nodes at depth $2i-2$ are marked with 1}   \} \\
&= 1- \sum_{l=1}^{M-1} \left[ 1- \underbrace{P\{ \text{one OR node  at depth $2i-2$ is marked with 0\}}}_{ \sum_{k=1}^K q_k y_{i-1,k}}  \right]^{l-1} \rho_l   \label{eq:weighted_sum} \\
&= 1- \sum_{l=1}^{M-1} \left[ 1- \sum_{k=1}^K q_k y_{i-1,k}  \right]^{l-1} \rho_l \\
&=1- \rho\left(1- \sum_{k=1}^K q_k y_{i-1,k}  \right)  \label{eq:rho}
\end{align}   
\end{subequations}
where in Eq. \eqref{eq:weighted_sum} the probability of having one OR node at depth $2i-2$ marked with $0$ is given by $\sum_{k=1}^K q_k y_{i-1,k}$. This corresponds to the probability of having a OR node at depth $2i-2$ and  of class $k$ marked with $0$, weighted by the probability of having a OR node of class $k$ as child. 
 Substituting Eq. \eqref{eq:rho} into Eq. \eqref{eq:lambda}, we obtain the following recursion
\begin{align}\label{eq:and_or_tree}
y_{i,k}  &=Ê \lambda_{k}\left( z_{i-1} \right)  \\
z_{i-1} &=  1- \rho\left(1- \sum_{k=1}^K q_k y_{i-1,k}  \right) \nonumber
\end{align}
with    $y_{0,k}=1, \forall k$, $q_0=1$ \footnote{In the above analysis we have assume a tree ensemble representation, which implies that the bipartite graph is loop-free, since loops introduce correlation in the evolution of the message error probabilities. This is true for large networks.}.
 Note that the error probability $y_{i,k}$ is  recursively derived assuming that   the OR nodes at depth $2i-2$ are actually roots of  trees with depth $2i-2$ and are independent from each others.    
Substituting  Eq. \eqref{SN_edge_degree_asympt} into Eq. \eqref{eq:and_or_tree}, we obtain
\begin{align}\label{eq:and_or_tree_asympt}
y_{i,k}  &=Ê \lambda_{k}\left( 1- \exp\left(- G \frac{\sum_k L_k \Lambda_k^{\prime}(1) }{\sum_k L_k} \sum_{k=1}^K q_k y_{i-1,k}  \right)  \right)
\end{align}
 Finally, let  $P_e(k,I)$ be the probability for the base station  of not correctly decoding the message sent from a BN of class $k$ when a iterative SIC technique  is adopted with a maximum of $I$ iterations. As already observed in Fig. \ref{fig:Tree}, the message (BN) of class $k$  can be decoded through any neighboring SN. It means that  $P_e(k,I)$  is computed as the probability that   the message cannot  be decoded through any edge at the $i$th iteration of the SIC algorithm, which means that 
 \begin{align} \label{eq:final_Pe}
P_e(k,I) = \sum_{l=0}^N  y_{I,k}^l \, \Lambda_{l,k}  
\end{align}

\begin{figure}[t]
\begin{center}
\includegraphics[width=0.24\linewidth,  draft=false]{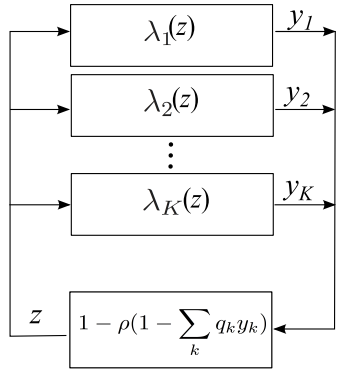}
\caption{Feedback loop associated to the iterative decoding probability described in Eq.  \eqref{eq:and_or_tree}. Each error probability $y_{i,k}$ is evaluated as a function of the $z_{i-1}$, while $z_{i-1}$ is evaluated as a function of a weighted sum of all $y_{i-1,k}$ probabilities. This recursion creates the feedback loop depicted in the figure. 
}\label{fig:Feedback_Loop}
\end{center}
\end{figure}

\subsection{Stability Conditions}
\label{sec:stability_conditions}
We are now interested in evaluating the conditions under which the SIC algorithm asymptotically ($i\rightarrow \infty)$ converges with zero failure probability. We first observe that the  iterative decoding process described in Eq. \eqref{eq:and_or_tree}  can be seen as a feedback loop (Fig. \ref{fig:Feedback_Loop}) of which we can study the  global stability, by deriving the conditions under which the system asymptotically  converges to null  error probability, i.e., to an equilibrium point $(\mat{y}^{\star},z^{\star}) = (\mat{0},0)$, for any   initial probability  $(\mat{y},z)$. The global stability can be guaranteed if  the error probability $z_i$ decreases at every decoding iteration, converging then to a zero error probability for $i\rightarrow \infty$.   

We first note that the  control equations that characterize the feedback system, as well as the iterative decoding probability of Eq. \eqref{eq:and_or_tree},  are 
\begin{align}
z & = 1- \rho(1-\sum_k q_k y_k) \nonumber \\
y_k & =  \lambda_k(z) \nonumber 
\end{align}
as shown in Fig. \ref{fig:Feedback_Loop}. By substituting Eq.  \eqref{SN_edge_degree_asympt} in the above equations, we obtain the following control equation for the feedback system
\begin{align}  
\label{eq:Global_stab}
 z &=  \underbrace{1- \exp\left(-  \frac{\sum_k L_k \Lambda_k^{\prime}(1) }{N} \sum_{k=1}^K q_k \lambda_{k}\left( z  \right)   \right)  }_{f(z)}
\end{align}
 and  we can guarantee global stability if $f(z)<z, \forall z$. This means that we should have
\begin{align} \label{eq:global_stability}
& 1- \exp\left(- G  \frac{\sum_k L_k \Lambda_k^{\prime}(1) }{\sum_k L_k} \sum_{k=1}^K q_k \lambda_{k}\left( z  \right)   \right)  <  z\,.  
\end{align}
In the following, we show how the global  stability can be exploited to solve the transmission optimization problem in prioritized MAC algorithms.

\subsection{Analytical Performance Validation}
\label{sec:Analyt_perf}
We now provide simulation results to validate the theoretical analysis provided above and study the performance of the random MAC transmission protocol in different settings.    We set the maximum number of iterations used in the burst decoding  algorithm to $I=100$.   For each simulated scenario, we  average the experienced utility function  over  $1000$ simulated loops.  We test the theoretical analysis for different transmission probabilities, provided in Table \ref{tab:Lambda}.

\begin{figure}[t]
\subfigure[BN edge-perspective]{
\includegraphics[width=0.44\textwidth,  draft=false]{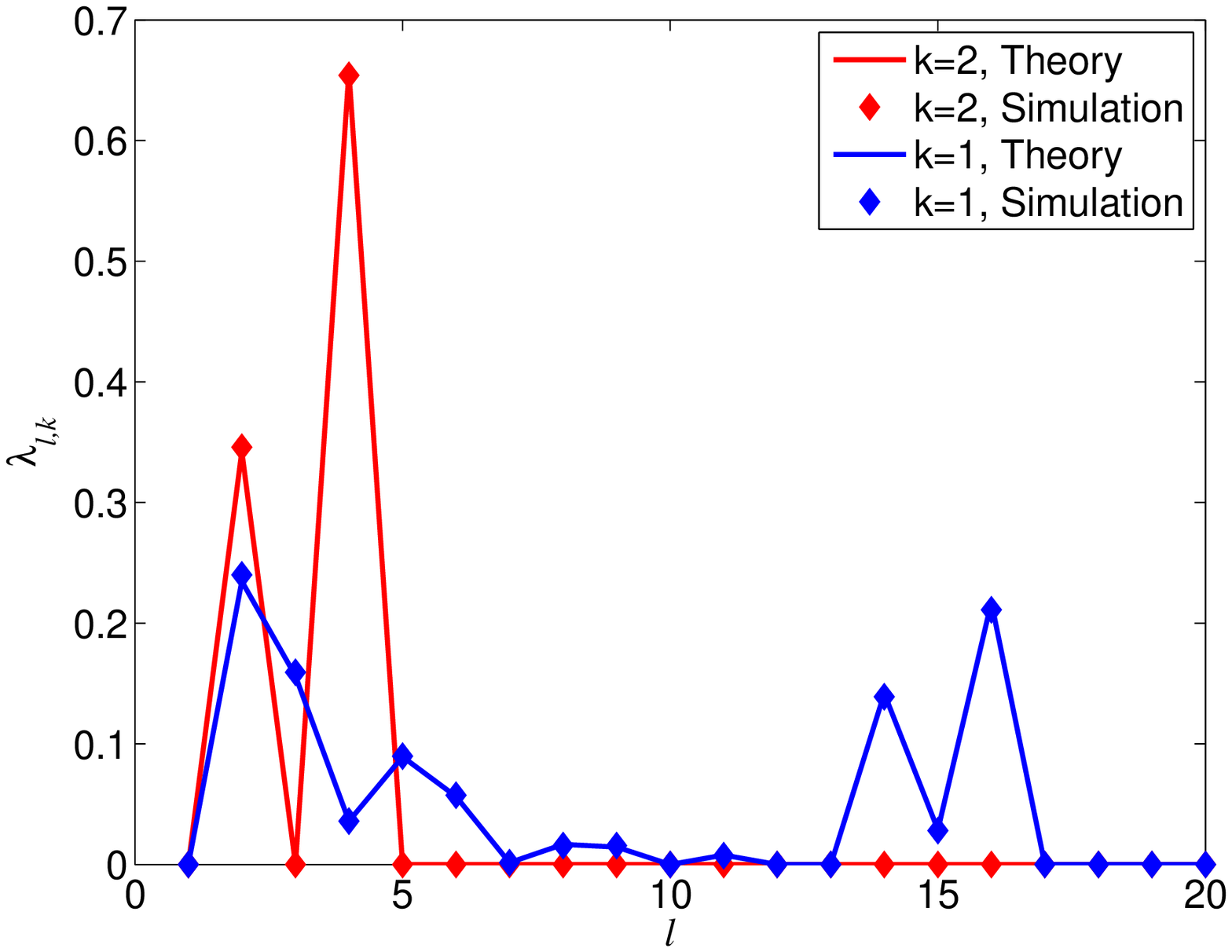}
\label{fig:} }\hfil
\subfigure[SN edge-perspective]{
\includegraphics[width=0.44\textwidth,  draft=false]{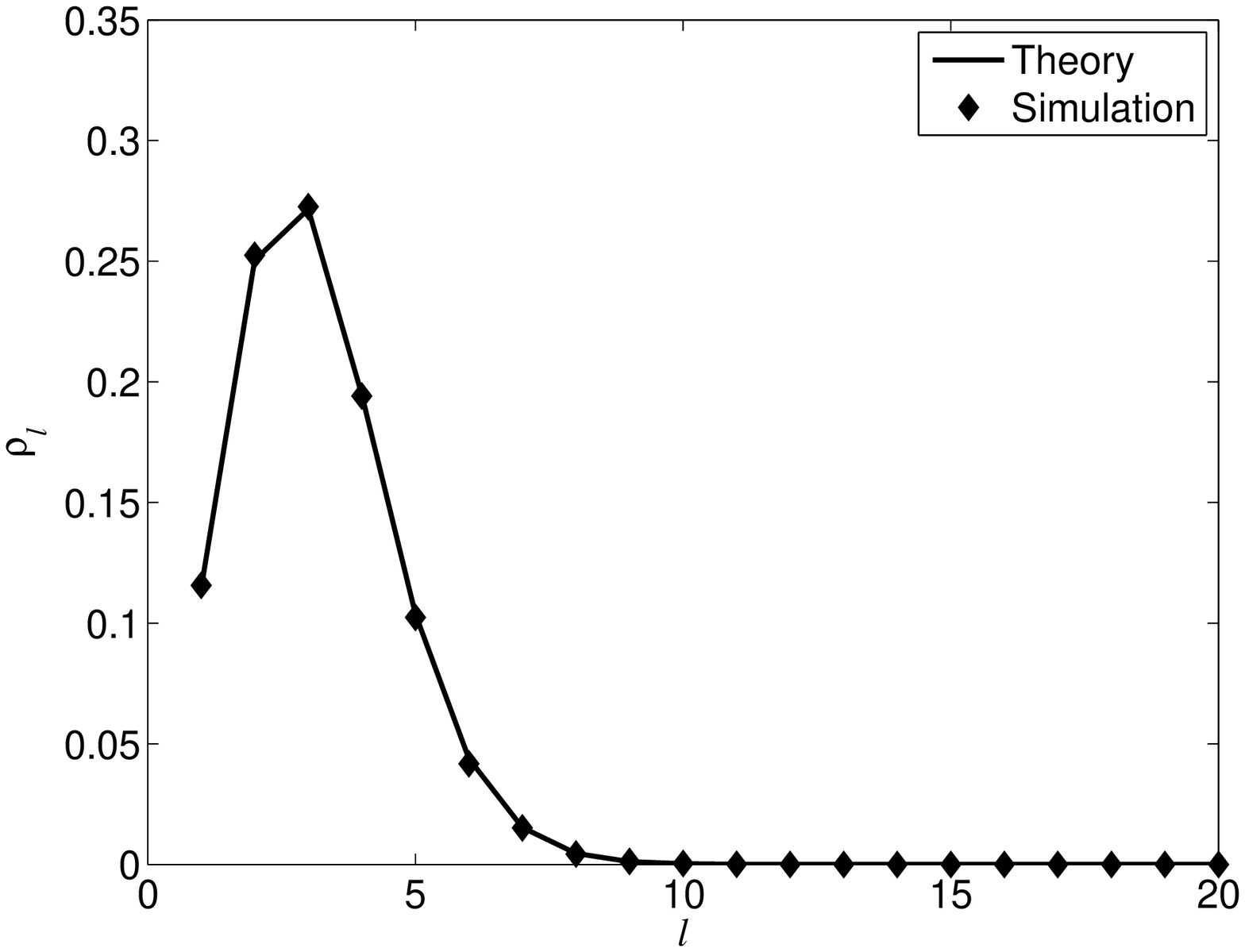}
\label{fig:} }\hfil
\caption{Comparison of the theoretical and simulated edge-perspective degree distributions for a system with $K=2$, $N=200$, $L_1=L_2$, and $\Lambda_1(x)=\Lambda^{\text{f}}(x)$ and $\Lambda_2(x)=\Lambda^{\text{a}}(x)$.} 
\label{fig:Statistical_check}
\end{figure}

  Fig.  \ref{fig:Statistical_check} first  depicts the theoretical and simulated edge-perspective degree distribution for a scenario   with $N=200$ transmission slots in a MAC frame,  $K=2$ source classes with the same number of burst nodes (i.e., $L_1=L_2$), and   with the following degree distributions $\Lambda_1(x)=\Lambda^{\text{f}}(x)$ and $\Lambda_2(x)=\Lambda^{\text{a}}(x)$ that have been shown to be effective in random MAC strategies \cite{Liva:J11}. We see that the theoretical performance is in agreement with the simulation results, which means that  the assumption of large networks used in the analysis still      holds in the case of finite size of the MAC frame ($N=200$), as observed   in other works \cite{Liva:J11}.

\begin{table}[t]
\centering
\caption{Considered polynomial distributions per class $\Lambda_k$.}\label{tab:Lambda}
\begin{tabular}{|c|c|c|}
\hline
  Index & Label &  Transmission Probability Distribution $\Lambda_k(x)$  \\
\hline\hline
  1 & $\Lambda^{\text{a}}(x) $ &  $0.5102x^2+0.4898x^4$\\
 \hline
  2 &   $\Lambda^{\text{b}}(x) $ & $0.5631x^2+ 0.0436x^3+0.3933x^5 $\\
 \hline
   3 &  $\Lambda^{\text{c}}(x) $ & $0.5465x^2+0.1623x^3+0.2912x^6$ \\
\hline
   4 &  $\Lambda^{\text{d}}(x) $ & $0.5x^2+0.28x^3+0.22x^8$\\
\hline
   5 &  $\Lambda^{\text{e}}(x) $ & $0.08x^3+0.14x^4+0.3x^5+0.17x^6+0.14x^7+ 0.17x^9$ \\
\hline
   6 &  $\Lambda^{\text{f}}(x) $ & $0.4977x^2+0.2207x^3+0.0381x^4+0.0756x^5+0.0398x^6+$ \\  
   & &$0.0009x^7+0.0088x^8+0.0068x^9+0.0030x^11+0.0429x^{14}+0.0081x^{15}+0.0576x^{16}$\\
\hline
\end{tabular}
\end{table}

\begin{figure}[t]
\begin{center}
\subfigure[Normalized Throughput for class $1$]{
\includegraphics[width=0.44\linewidth,  draft=false]{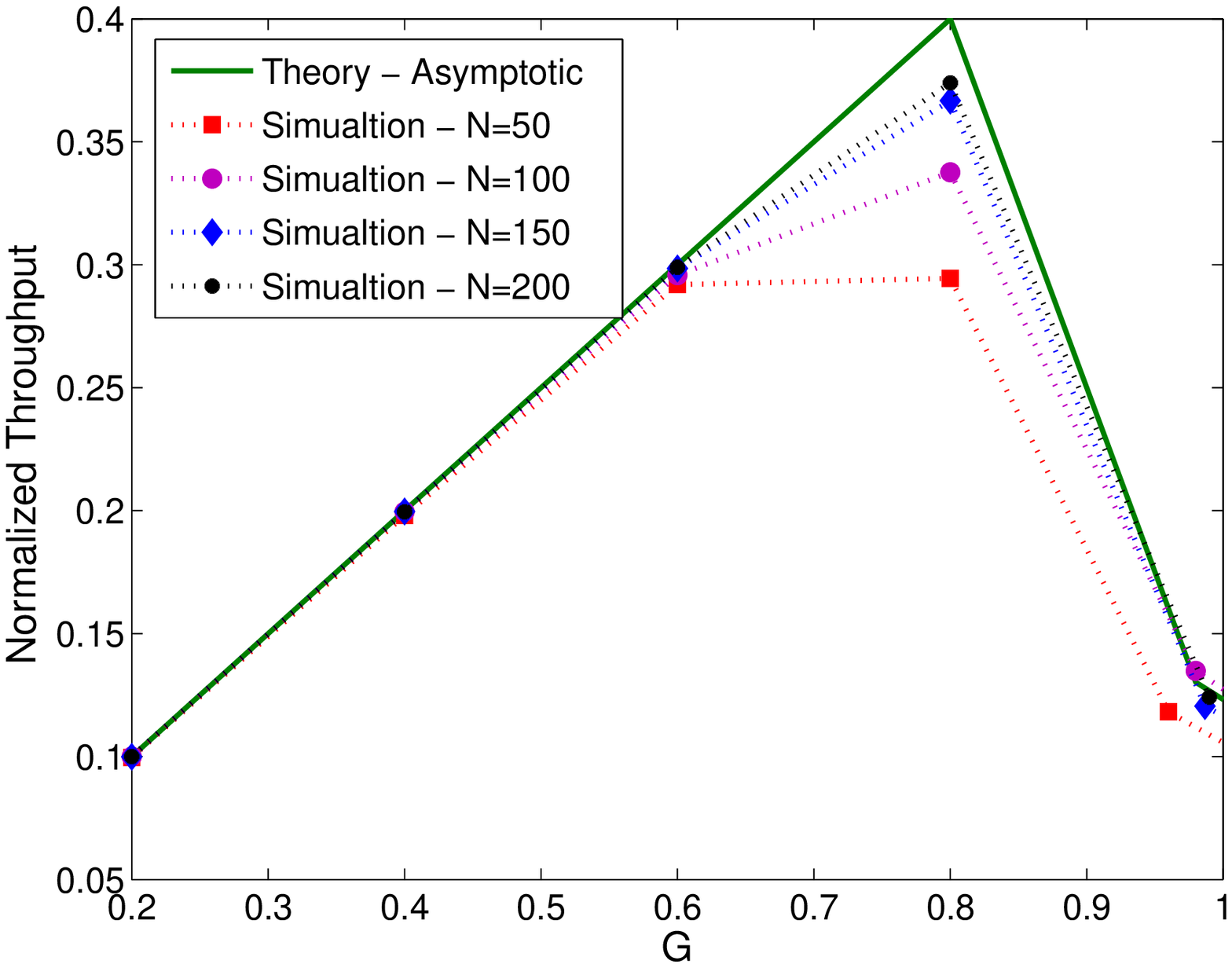}\label{fig:normalized_thr_LambdaOPT_Lconst}}
\subfigure[Mean Utility Function]{
\includegraphics[width=0.44\linewidth,  draft=false]{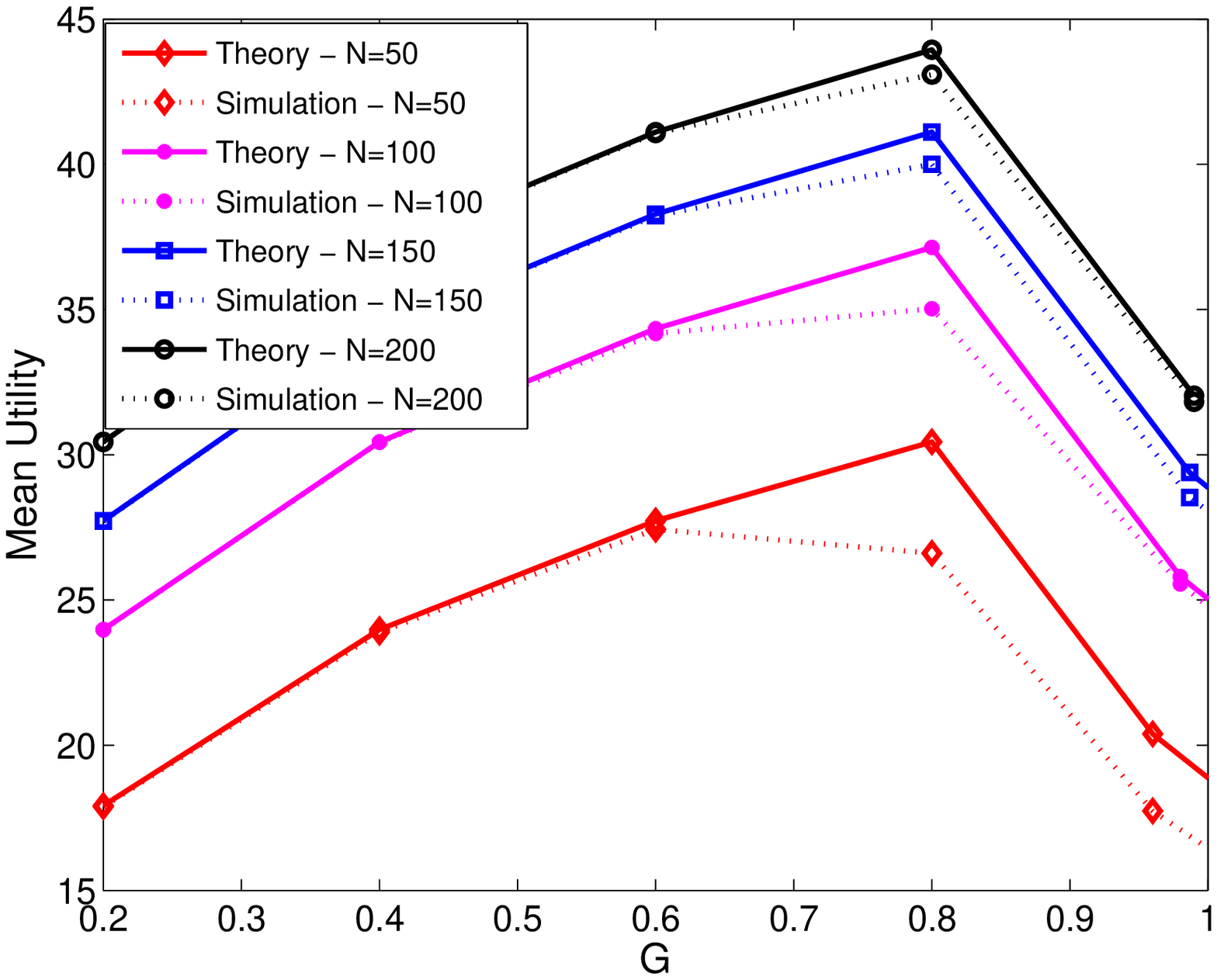}\label{fig:Utility_LambdaOPT_Luguale_Nvarying}}
\caption{Heterogenous system performance vs. network load $G$  for the case of $K=2$, $w_1=0.7, $ and $w_2=0.3$,  $L_1=L_2$, $\Lambda_1(x)= \Lambda^{\text{d}}(x)$ and $\Lambda_2(x)= \Lambda^{\text{a}}(x)$. Different numbers of transmission slots $N$ are considered.}\label{fig:p025_Lconstant_Nvarying}
\end{center}
\end{figure}

We further  provide a comparison between simulations and theoretical results in terms of both normalized throughput and utility function.   We  consider an illustrative scenario with  two classes with priority given by  $w_1=0.7$ and $w_2=0.3$ (this   means that class $1$ is more important than class $2$), with the same number of burst nodes    $L_1=L_2$, and  two different distributions $\Lambda_1(x)= \Lambda^{\text{d}}(x)$ and $\Lambda_2(x)= \Lambda^{\text{a}}(x)$.  In Fig. \ref{fig:p025_Lconstant_Nvarying}, both  the average of the  normalized throughput    and the overall utility function are  depicted as a function of the traffic load $G$ for $1000$ simulation runs, for different size of $N$ of the MAC frame. The traffic load $G=\sum_k L_k/N$ varies with the number of burst slots $L_1=L_2$, for a fixed $N$.  The normalized throughput   is evaluated as $G(1-P_e(k,I))$, and we only show the throughput for class $1$ since the results are mostly similar for all classes. We see that   there is again a good match between theoretical and simulated results for all $N$ larger than $50$.  When $N=50$, the error in considering an asymptotic behavior is non negligible, but yet the utility function behavior is accurate. In particular, when $N=50$, both theoretical and simulation values of the mean utility increases  with $G$ for $G<0.6$, have a peak between $[0.6,0.8]$, and   decrease beyond a traffic value of $0.8$. We have obtained similar trends in other experiments for different scenarios, and all confirm the validity of the analysis for large MAC frames and the possibility to characterize the optimal traffic load, even with small MAC frames.

\begin{figure}[t]
\begin{center}
\subfigure[Error Probability]{
\includegraphics[width=0.44\linewidth,  draft=false]{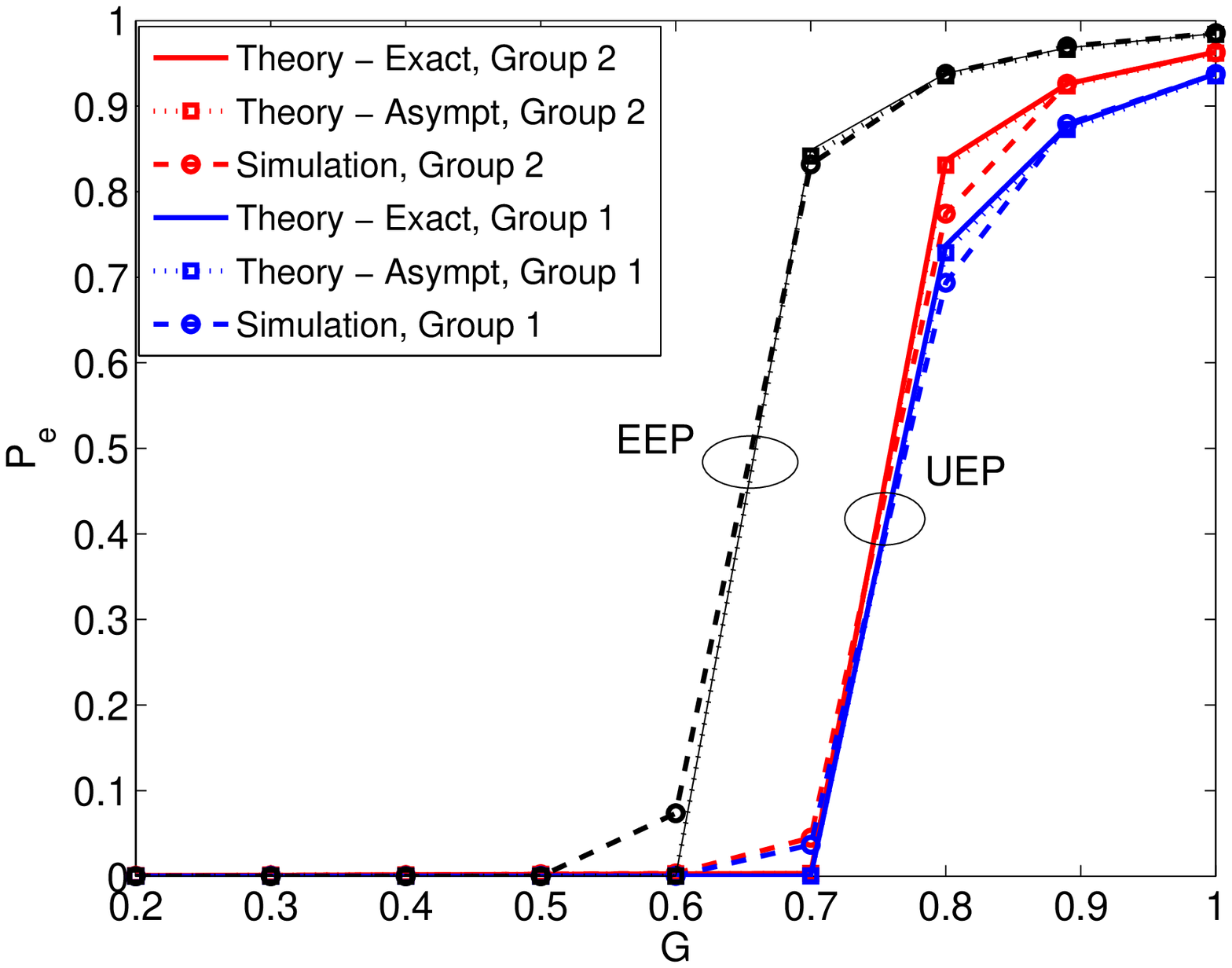}\label{fig:Pe_p025_Lconstant}}
\subfigure[Mean Utility Function]{
\includegraphics[width=0.44\linewidth,  draft=false]{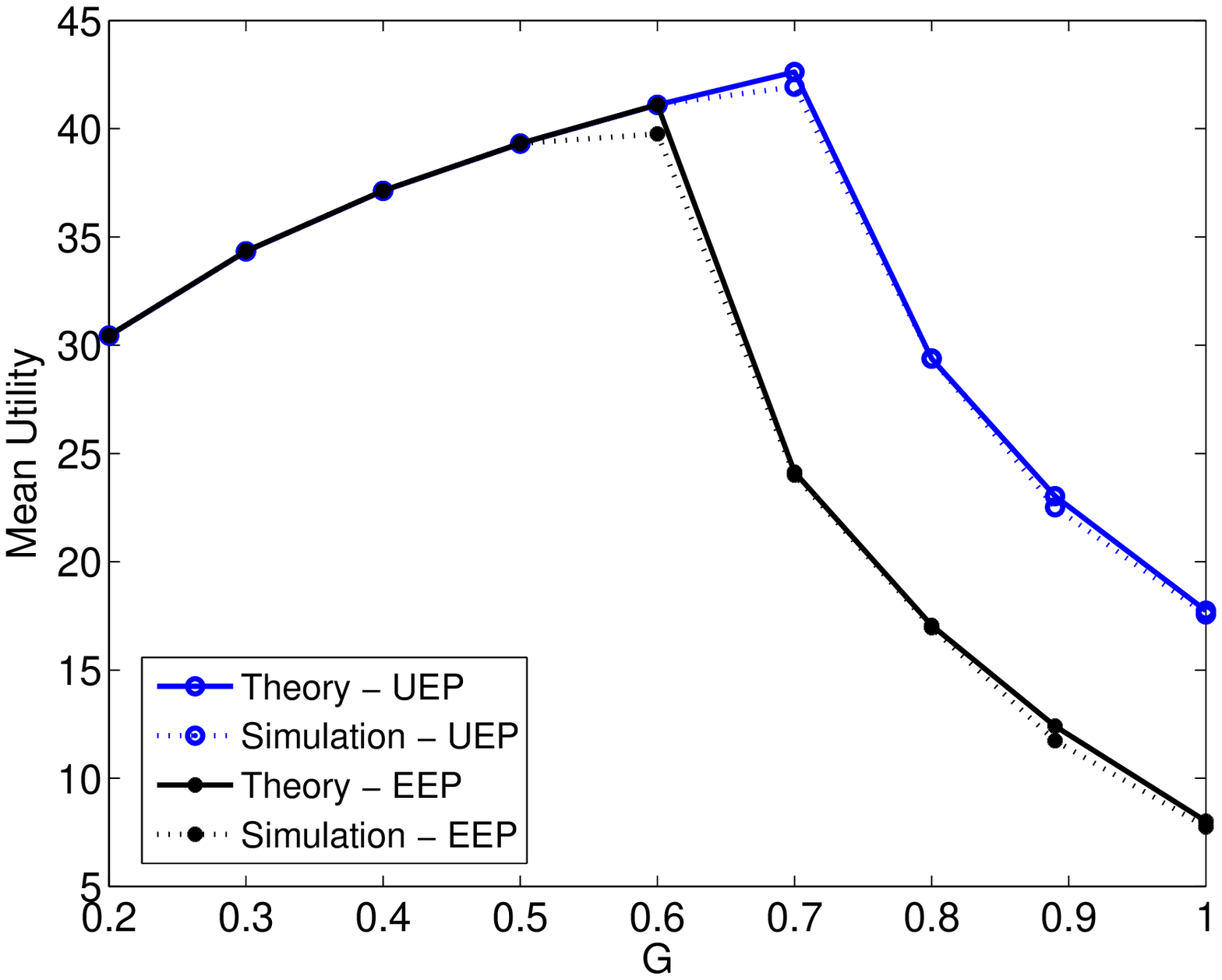}\label{fig:Mean_Utility_p025_Lconstant}}
\caption{System performance vs. network load $G$  for the case of $N=200$, $K=2$, $w_1=0.7, $ and $w_2=0.3$, $L_1=L_2$. For the UEP case $\Lambda_1(x)=  \Lambda^{e}(x)$, and $\Lambda_2(x)= \Lambda^{b}(x)$
 For the EEP case, $\Lambda_1(x)=\Lambda_2(x)=\Lambda^{e}(x)$.}\label{fig:p025_Lconstant}
\end{center}
\end{figure}

We now illustrate the performance of prioritized  transmission schemes  for sources with different importance.    In Fig. \ref{fig:p025_Lconstant},      the message error rate and the overall utility function are  given  as  functions of the traffic load $G$ for a sample scenario with $N=200$ transmission slots, two classes with the same number of slots ($L_1=L_2$)  but with  different priorities ($w_1=0.7$ and $w_2=0.3$). The traffic load $G=\sum_K L_K/N$ again varies with the number of burst nodes in each class. In order to illustrate the benefits of prioritized transmission when sources are heterogeneous, we compare the performance of equal and unequal transmission probabilities for different classes. For the  equal transmission strategy, denoted by EEP,  we consider $\Lambda_1(x)=\Lambda_2(x)=\Lambda^{e}(x)$; while for the case of unequal transmission strategy, denoted by UEP,   we consider  different distributions $\Lambda_1(x) = \Lambda^{e}(x)$ and $\Lambda_2(x)= \Lambda^{b}(x)$, which correspond to different transmission probabilities (namely higher replication rate for the most important classes).  We first observe that  theoretical results again match  the simulations results computed over $1000$ simulation runs. The results also show the benefit of  prioritized transmission policies when source have different priorities. In  Fig. \ref{fig:Pe_p025_Lconstant}, we see that the message error probability per class is higher  for the   EEP strategy than for prioritized strategy under consideration.  We also observe that a EEP strategy leads to a waterfall effect in proximity of a threshold value of $G = 0.6$, while the UEP strategy has a larger threshold value of $G = 0.7$. A larger threshold value implies a larger throughput, since  a larger number of sources   actively transmit messages within a given MAC. This gain achieved by the UEP strategy is due to the change   in the transmission protocol for different  classes: in our example, the class $2$ adopts a replication rate following   $\Lambda_2(x)$ rather than $\Lambda_1(x)$, with a maximum node degree of $5$ rather than $9$. This reduces the overall number of messages sent in a  MAC frame, hence also the load   of the network. This reduced replication rate  mainly affects class $2$ rather than class $1$, which is more important in our example. This leads to an overall utility function that reaches a maximum of  $43.2 $ dB for the UEP strategy, as opposed to the maximum of $40.8 $ dB of the EEP strategy, see Fig. \ref{fig:Mean_Utility_p025_Lconstant}.

\begin{figure}[t]
\begin{center}
\includegraphics[width=0.44\linewidth,  draft=false]{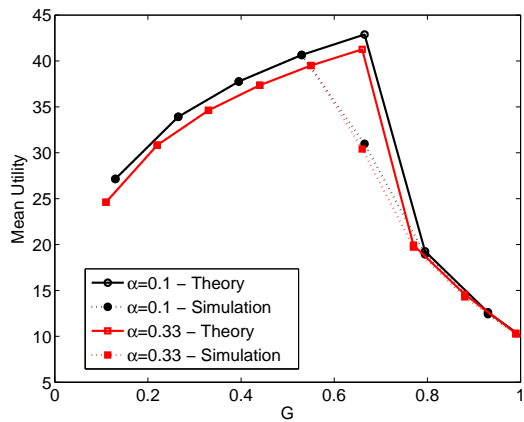}\caption{System performance vs. network load $G$  for the case of UEP transmission, for  $N=200$, $K=2$, $w_1=0.7, $ and $w_2=0.3$, $L_2= \alpha  L_1$, $\Lambda_1(x)=\Lambda^{e}(x)$, and $\Lambda_2(x)=\Lambda^{b}(x)$.  }\label{fig:Example_DifferentRatio_3}
\end{center}
\end{figure}

Another example of prioritized transmission is  provided in Fig. \ref{fig:Example_DifferentRatio_3}, where the same scenario of Fig. \ref{fig:p025_Lconstant}  is considered, but with  a different number of burst nodes  in each class, i.e.,  $L_2=\alpha L_1$, with $\alpha=0.1$ and $\alpha=0.33$. We see again that the theoretical and the simulation results are generally in accordance, especially in the low and high traffic regions. We further see that the number of burst nodes per class affects the performance of the prioritized transmission solution. In particular, reducing the number of burst in the lower importance class improves the overall system performance.  Finally, we    observe  that   theoretical study  is  an upper bound of the simulated performance and the theoretical value of $G^{\star}$ is an upper bound of the simulated value of $G^{\star}$, where $G^{\star}$ the threshold value beyond which  the error probability rapidly reaches $1$ and a waterfall effect is experienced. 

Overall, the above results illustrate that prioritized transmission is beneficial when sources have different importance, and that the system performance is dependent on both the replication rates (or the transmission probabilities) and the number of burst nodes.  In the following, we show how this  theoretical study of MAC protocol strategies can be adopted in practical resource allocation optimization problems.

\section{Prioritized Random Access   Optimization}
\label{sec:SolvingMethod}
 Based on the above theoretical analysis, we formulate now a transmission policy optimization aimed at finding the best  $(\Lambda_k(x), L_k)$ per class, such that the network resources are not   under-utilized  but not over-utilized either.   More in details, we choose both the transmission probability and the source rate for each class, such that the overall system performance is maximized.  The optimization, carried out by the base station, finds the best tradeoff between throughput and message decoding probability for the all sources. 

\subsection{Problem Formulation}
Let denote by $\mat{\Lambda} = [\Lambda_{1}(x), \Lambda_{2}(x), \ldots, \Lambda_{K}(x)]$ the transmission policy vector with the transmission probability distributions for each class.   By varying  $\mat{\Lambda}$    and the number of   bursts  $L_k$ for  source  class $k$, we can actually modify   the expected received throughput for each class.  As the throughput   corresponds to an  utility function score for each class, as shown in Eq. \eqref{eq_exp_Dist}, we can modify the overall system performance by tuning $\mat{\Lambda}$ and $\{L_k\}$.  In particular, we can optimize the overall utility of a system with heterogenous sources by maximizing  the throughput of  most important classes and sacrificing the throughput of  least important ones.  This leads to prioritized transmission strategies that are necessary for systems with sources that have different utility functions. Hence, instead of maximizing   the overall throughput, we optimize  the best  transmission strategy  $(\mat{\Lambda},\mat{L})$ and select   the one that maximizes the overall utility function, evaluated as follows
\begin{align} \label{eq:pf_1}
(\mat{\Lambda}^{\star} ,\mat{L}^{\star} ) &= \arg\max_{(\mat{\Lambda},\mat{L})}\overline{U}(\mat{\Lambda},\mat{L})  \\ \nonumber
& = \arg\max_{(\mat{\Lambda},\mat{L})} \left\{
 \sum_{R_1=0}^{L_1} \sum_{R_2=0}^{L_2} \ldots \sum_{R_K=0}^{L_K}  U(R_1, R_2, \ldots, R_K) P_r^{(I)}(R_1, R_1, \ldots, R_K; \mat{\Lambda},\mat{L})   \right\}
\end{align}
where $U(R_1, R_1, \ldots, R_K)$ is the system utility function defined in Eq.  \eqref{eq_exp_Dist} while 
 $P_r^{(I)}(R_1, R_1, \ldots, R_K; \mat{\Lambda},\mat{L})$ is the probability of correctly receiving $R_1, R_2, \ldots, R_K$ messages for sources of class  $1$, $2, \ldots, K$, respectively. This corresponds to the probability of receiving  $R_1, R_2, \ldots, R_K$ messages either with no collisions or   messages with collisions that can be resolved  with the iterative message recovering strategies (i.e., SIC) after a maximum number of  $I$ iterations.   
 

Since the transmission processes are handled independently by all sources, we can derive the probability of correctly receiving the the messages from different sources as follows 
 \begin{align} 
P_r^{(I)}(R_1, R_1, \ldots, R_K; \mat{\Lambda},\mat{L}) &= \prod_{k=1}^K P_{k}^{(I)}(R_k; \Lambda_{k}(x), L_k) \\ \nonumber  
&= \prod_{k=1}^K
  {L_k \choose R_k}   [1-P_e(k,I)]^{R_k} P_e(k,I)^{L_k-R_k} 
\end{align}
where    $ P_{k}^{(I)}(R_k; \Lambda_{k}(x), L_k)$ is the probability for a source of class  $k$ to recover $R_k$  out of $L_k$  messages after  the $I$-th IC iteration,  and   $1- P_e(k,I)$ is the probability for the base station to correctly receive  the   message sent from   a source of class $k$. 
We can then express the expected distortion as 
\begin{align}\label{eq:exp_dist2}
\overline{U}(\mat{\Lambda},\mat{L}) = \sum_{R_1=0}^{L_1} \sum_{R_2=0}^{L_2} \ldots \sum_{R_K=0}^{L_K}  \sum_{k=1}^K w_k  U_k(R_k) \prod_{R_K=0}^{L_K} {L_k \choose R_k}   [1-P_e(k,I)]^{R_k} P_e(k,I)^{L_k-R_k}\,. 
\end{align}
and the problem formulation to be solved becomes 
  \begin{subequations}\label{final_pf}
\begin{align}  
(\mat{\Lambda}^{\star},\mat{L}^{\star})  : \arg\max_{\mat{\Lambda},\mat{L}} \ &\overline{U}(\mat{\Lambda},\mat{L})  
\\  \label{final_pf_aa}
\text{s.t. } &\Lambda_k(x)\leq \Lambda_{k+1}(x) &  \forall x\in[0,1], \forall k
  \end{align}
\end{subequations}
where the priority  constraint in Eq. \eqref{final_pf_aa}  permits to reduce the search space. In particular, we constraint the optimization to  an unequal recovery probability among classes such that sources from more important classes have a larger probability of correctly transmitting their messages compared to lower important sources. This translates in imposing  $P_e(k,i)   \leq P_e(k+1,i)$  for class $k$ more important than class $k+1$. From Eq. \eqref{eq:final_Pe},  the priority  condition  can be generalized as $\Lambda_k(y_{i-1,k})\leq \Lambda_{k+1}(y_{i-1,k+1})$. The optimization problem in Eq. \eqref{final_pf} results in maximizing the  overall system utility.

\subsection{Approximated Solution}
The optimization problem in Eq. \eqref{final_pf} might not be easily solved with conventional optimization frameworks. The expected utility function is evaluated as a weighted sum of binomial  distributions, each of them having  a probability   $P_e(k,I)$.   Although $P_e(k,I)$ can be considered as a sigmoid function to simplify the formulation,   there is no a convenient  optimization framework that is  able to address the above problem jointly for both $\mat{\Lambda}$ and $\mat{L}$ variables,     to the best of our knowledge. 
Thus, we propose a solving method that  exploits an intrinsic property  of the coded slotted ALOHA:  the message error probability usually follows a waterfall effect~\cite{richardson:B08},   having an error probability approaching $0$ for traffic network $G$ lower than a given threshold $G^{\star}$, and rapidly approaching $1$ beyond $G^{\star}$.  The threshold value  $G^{\star}$ is usually defined as the value  that is the limit of the region where  the    condition of  stability  hold.

In the following, we approximate the message error probability $P_e(k,I)$  to $0$ when   the stability condition is respected, i.e., when  the convergency of the iterative message decoding algorithm is assumed. 
By imposing the global stability condition of Eq. \eqref{eq:global_stability}, we have the following instance of the optimization problem: 
  \begin{subequations} \label{eq:PF_approx_global}
\begin{align}
 (\mat{\Lambda}^{\star},\mat{L}^{\star})  : \arg\max_{\mat{\Lambda},\mat{L}} \ &\sum_{k=1}^K w_k U(L_k)
\\     \label{eq:PF_approx_C1}
 \text{s.t. } &\Lambda_k(x)\leq \Lambda_{k+1}(x) &  \forall x\in[0,1], \forall k
\\   &   \label{eq:PF_approx_C2}
   \exp\left(- G  \frac{\sum_k L_k \Lambda_k^{\prime}(1) }{\sum_k L_k} \sum_{k=1}^K q_k \lambda_{k}\left( x \right)   \right)  >1- x &\forall x\in[0,1]\,. 
  \end{align}
\end{subequations}
  where we have assumed a null error probability if the constraint in Eq. \eqref{eq:PF_approx_C2} is met for all classes, i.e., if the conditions for global stability are met. This means that  the $L_k$ messages   sent from sources of  class $k$ are correctly  received. The experienced utility function in each class is therefor     $U_k(L_k)$, and the global utility is the weighted sum of class utilities.   

In the following, we denote by \emph{ON region} the set of  pairs of  $(\mat{\Lambda},\mat{L})$ such that the stability condition   is respected.  
The above  optimization problem    permits  to select the ON region of the system and to seek for a solution within the region where all packets are decoded. This instance of our optimization problem in Eq.  \eqref{final_pf} offers simpler solution to the selection of the best transmission strategy.

\begin{algorithm}[t]                      
\caption{Prioritized Random Access Protocol Optimization}          
\label{alg_OPT}
\begin{algorithmic}[1]                    
\STATE \textnormal{{\bf \textit{step 1)}}:  Define the set $\mathcal{L}_{\text{ON}}$ defined as the set of pairs  $(\mat{\Lambda},\mat{L})$ in the ON region:   }
$$
\mathcal{L}_{\text{ON}}: \left\{(\mat{\Lambda},\mat{L}) \text{ s.t.   the constraint in   Eq. \eqref{eq:PF_approx_C2} is verified} \right\}$$
\STATE \textnormal{{\bf \textit{step 2)}}: Optimize the utility function within the ON region as follows 
\begin{align}
(\mat{\Lambda}^{\star},\mat{L}^{\star})  : \arg\max_{(\mat{\Lambda},\mat{L}) \in \mathcal{L}_{\text{ON}}} \ & \sum_{k=1}^K w_k \, U(L_k) \nonumber \\ \nonumber
\text{s.t. } &\Lambda_k(x)\leq \Lambda_{k+1}(x) &  \forall x\in[0,1], \forall k
\end{align}
}  
 \end{algorithmic}
\end{algorithm}

The optimization problem can be easily solved in two steps, as described in   Algorithm \ref{alg_OPT}.   The first step (Step $1$) defines the boundaries of the ON region by simply solving Eq. \eqref{eq:PF_approx_C2}, for the    global stability. Then, finding the best pair $(\mat{\Lambda}^{\star},\mat{L}^{\star}) $ in the ON region reduces to solving the optimization in Step $2$.  This optimization has  an objective function that has the form of  a weighted sum of utility functions  subject to $K$ affine constraints. Thus, it can be easily solved   for concave or linear $U(L_k)$ by concave or linear programming optimization, or by more general gradient-based optimization methods   for more general    non-decreasing  utility functions $U(L_k)$.

\subsection{Simulation Results}
We now provide simulation results to  study the performance of the prioritized random MAC transmission protocol in different settings.    We set the maximum number of iterations used in the burst decoding  algorithm to $I=100$.    For each simulated scenario, we  average the experienced utility function  over  $1000$ simulated loops. As utility function, we consider $U_k(R_k) = \log(R_k), \forall k$, which resembles a typical image quality metric.  In Section \ref{sec:Analyt_perf}, we have 
shown  that the  theoretical study of Section \ref{sec:Collision_Recovery_Probability} is  an upper bound of the simulated performance and the theoretical value of $G^{\star}$ is an upper bound of the simulated value of $G^{\star}$, where $G^{\star}$ is the end of the ON region where the error probability is negligible. While the asymptotic theoretical bound is surely  good to design reliable replication rates $\Lambda(x)$,   it might  lead to approximate solutions for   the optimization of the resource allocation $\mat{L}$  in finite MAC size. In the following, first  we show the main limitations of the optimization method based  purely on the theoretical study. Then,  we describe how these bounds combined with well-chosen heuristics can be adopted to jointly optimize $(\mat{\Lambda},\mat{L})$ for selecting effective prioritized transmission strategies and we show that the proposed approximated algorithm still achieves performance that is close to the optimal one. 

%

\begin{figure}[t]
\begin{center}
\includegraphics[width=0.44\linewidth,  draft=false]{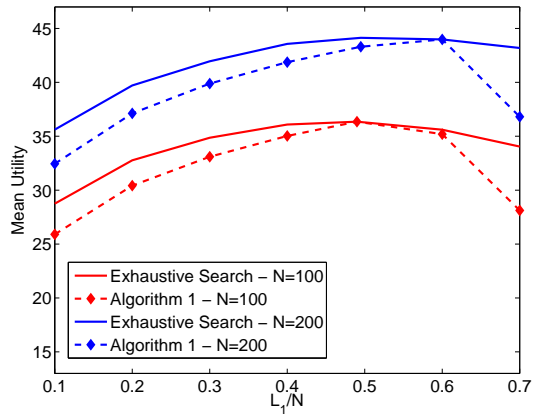} 
\caption{ Optimized system performance vs $L_1$ for the case of  $\Lambda_1(x)=\Lambda_2(x)= \Lambda^{e}(x)$.  The best value of $L_2$ is optimized for each  value $L_1$. }\label{fig:Optimized_L2_given_L1__Lambdaconst}
\end{center}
\end{figure}

We first study the optimization of the number of burst nodes per class. 
In Fig. \ref{fig:Optimized_L2_given_L1__Lambdaconst}, we show the global utility function as a function of the number of burst nodes in class $1$ (namely,  $L_1$ normalized by $N$),  for the case of two source classes, with  different importance  $w_1=0.7$, $w_2=0.3$, identical transmission probability distributions $\Lambda_1(x)=\Lambda_2(x)$, and $N=100 $ or $200$. For each value of $L_1$, we evaluate the best  value of $L_2$ and plot the corresponding global utility   score. The Algorithm \ref{alg_OPT} is considered for the optimization, where    the   global stability condition  has been imposed  by   Eq. \eqref{eq:PF_approx_C2} in the first step of the optimization algorithm. These results are compared with simulations results where the optimal $L_2$ for each $L_1$ is found by exhaustive search.    For both $N=100$ and $N=200$, 
 we notice that the global stability condition imposes a  stringent condition, leading to a tight  bound of the ON region compared to the simulated one. This leads to a mean utility score that is  almost as good as the optimal one. However, we notice that the model is not highly reliable for large values of $L_1$. For example, for $L_1/N =0.7$ in both  cases of $N=100$ or $200$ slots, the theoretical optimization leads to a substantial drop in the utility function.  This drop is due to the mismatch between theoretical and simulation results around   $G^{\star}$, as discussed in Section \ref{sec:Analyt_perf}.

\begin{figure}[t]
\begin{center}
\includegraphics[width=0.44\linewidth,  draft=false]{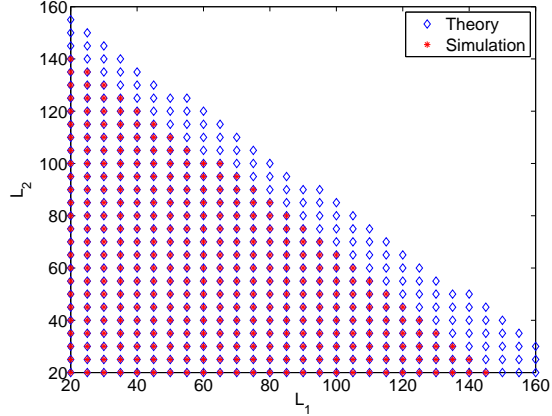} 
\caption{Example of ON region for a system with two source classes, $N=200$, $\Lambda_1(x)=\Lambda^{e}(x)$, and $\Lambda_2(x)= \Lambda^{a}(x)$. }\label{fig:ON_Region}
\end{center}
\end{figure}

 We now better study the effect of the mismatch between theoretical stability conditions and actual ones in the optimization algorithm. We recall that the ON region is the one that satisfies the global stability in Eq. \eqref{eq:PF_approx_C2}.  We show in Fig. \ref{fig:ON_Region}  the ON region  for a case  with  $K=2$ classes,  $N=200$ transmission slots, and the transmission probability distributions are $\Lambda_1(x)=\Lambda^{e}(x)$, and $\Lambda_2(x)= \Lambda^{a}(x)$. The ON region boundaries are derived
as a function of the number of burst nodes (or equivalently the traffic)    both from simulations and from the  theoretical analysis through the global stability condition of Eq. \eqref{eq:Global_stab}.  In the simulations we evaluate the ON region as the one where the decoding error probability after SIC is lower than $10^{-4}$.
As expected, the theory gives an ON region (in blu diamonds) that is more extended than the actual one (red points), since the theoretical ON region   gives an upper bound  on the value of traffic $G^{\star}$ that represents the transition between ON and OFF regions of the SIC algorithm. Unfortunately,  the best $\mat{L}$  derived from  Algorithm \ref{alg_OPT}   approaches  the boundary of the ON region defined by theoretical stability conditions, which is exactly the unsafe region where the theory does not necessary match the actual behavior of the system.  This means that for most cases, the optimization of Algorithm \ref{alg_OPT} would select a   transmission strategy such that the network is actually overloaded, which results in  a poor mean utility function. Based on these observations,   we can actually overcome the main limitation  of the theoretical study with the effective following heuristics.

The key concept is that we would like the system to work \emph{almost} at the boundaries of the ON region, but not \emph{exactely} at the boundaries. A first solution is to use the theoretical study to evaluate the boundaries, and then run   simulations in a neighborhood of the boundaries to predict the actual ON region. This method is however   not always   feasible because of the computational complexity in simulating the considered scenario.  
However, for all the scenarios considered in our work, we empirically observe   that the theoretical ON region extends beyond the actual one by about $0.1 G$. Thus, a good heuristic solution, called Algorithm 2, consists in first evaluating   the theoretical bound on the ON region and then translating   into a ``safe boundary" by reducing the boundaries by $10\%$. Finally, we can  seek  for the best  $(\mat{\Lambda},\mat{L})$ within the safe boundary region only, where the decoding error probability is actually zero. It is worth noting that the best $(\mat{\Lambda},\mat{L})$ is selected as the best resource allocation  
$\mat{L}$ such that the utility function is maximized and $(\mat{\Lambda},\mat{L})$ is on the   \emph{actual} ON region boundaries.  Note that it is much better to work with a traffic load that is slightly lower than the optimal one $G^{\star}$, rather than working at a traffic load slightly larger than $G^{\star}$. Working beyond the actual ON boundaries leads to  a state of error probability that is quickly approaching $1$, such that the achieved throughput  can quickly fall to   zero.  In the following, we illustrate this statement by comparing the performance of  Algorithm \ref{alg_OPT} and   Algorithm 2. 

%

\begin{figure}[t]
\begin{center}
\subfigure[Mean utility vs $L_1$ for $\Lambda_1(x)=\Lambda_2(x)= \Lambda^c(x)$.]{
\includegraphics[width=0.44\linewidth,  draft=false]{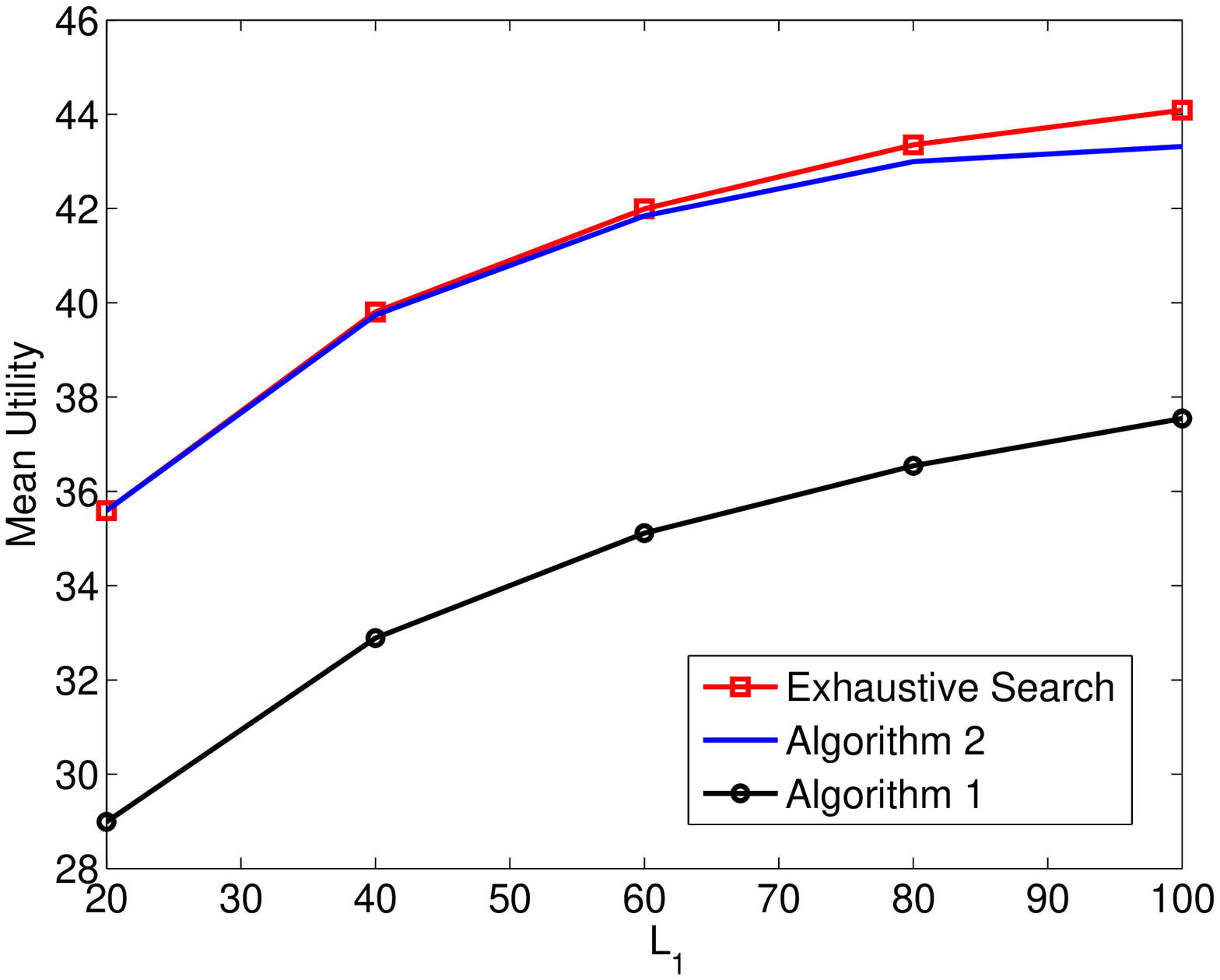}\label{fig:L1varying_N200_Safearea}}
\subfigure[Mean utility vs $\Lambda_1(x)$ when $L_1, L_2,$ and $\Lambda_2(x)$ are optimized.]{
\includegraphics[width=0.44\linewidth,  draft=false]{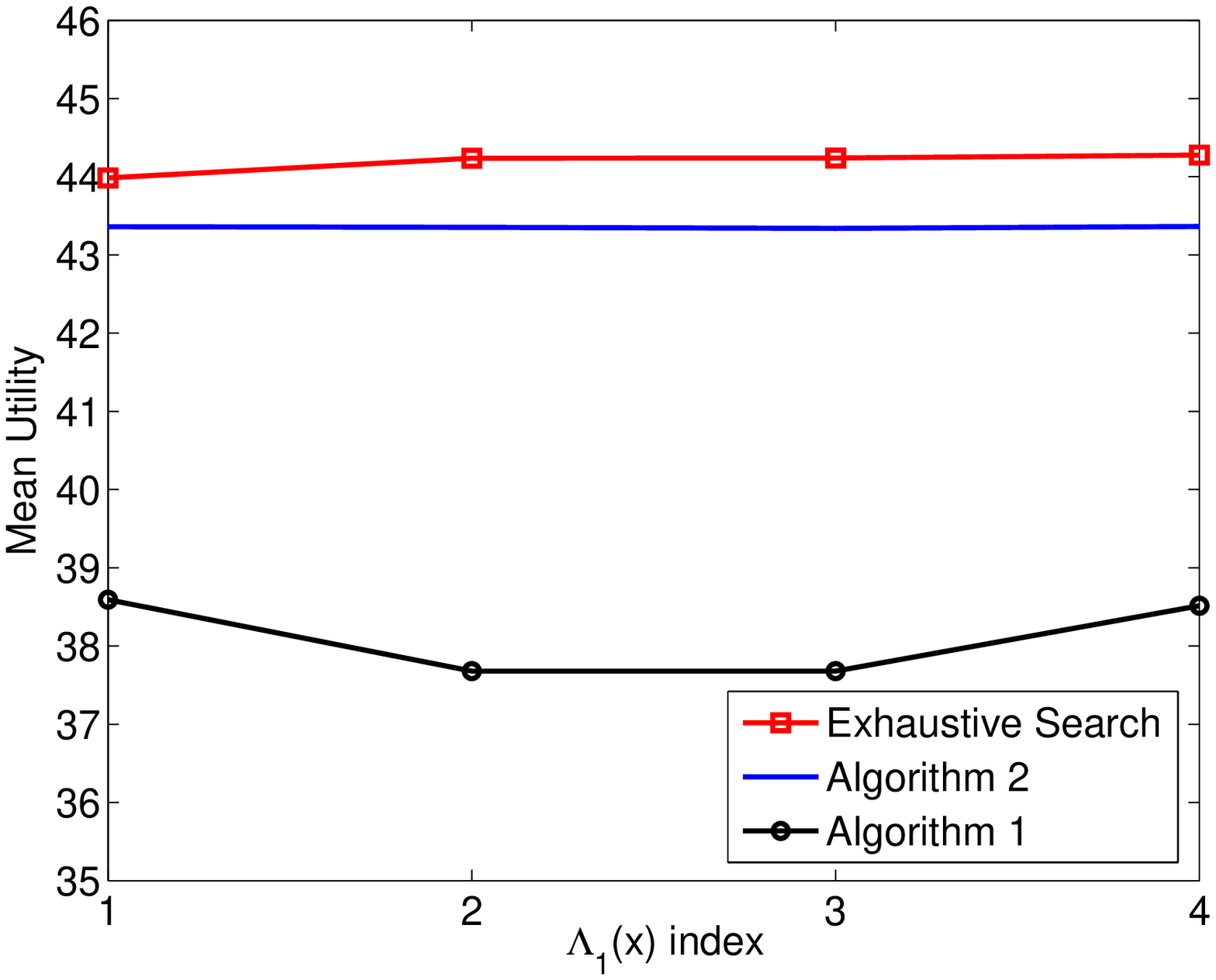}\label{fig:Different_Lambda_N200_w07}}
\caption{Optimization performance for Algorithm 1 and Algorithm 2   for a system with $N=200$, and source importance given as  $(w_1,w_2)=(0.7,0.3)$.}\label{fig:ONreg:comparison}
\end{center}
\end{figure}

  We propose now experiments where we compare the performance of Algorithm 1 and Algorithm 2.   To evaluate the set of pairs $(\mat{L},\mat{\Lambda})$ that satisfies the   stability constraints in the first step of both Algorithms,    Eq. \eqref{eq:PF_approx_C2} can be solved   for example by differential evolution \cite{DE:J97} as  shown in \cite{Liva:J11}. The best polynomial distribution can also be evaluated by numerical analysis by fixing a maximal degree for the BNs \cite{JakoveticBVC14a}.  However, for the sake of simplicity, in the following we consider the polynomial distributions  $\Lambda_k(x)$   derived   from \cite{Liva:J11} and provided in Table \ref{tab:Lambda}, which   have been   optimized to maximize the traffic threshold $G^{\star}$.  Any distribution in the table is a candidate for being assigned as  transmission strategy to a given source class.   However, our optimization   can   be applied to any other sets of polynomial distributions that satisfies   Eq. \eqref{eq:PF_approx_C2}. We compare the performance of both algorithms to an exhaustive search through all $(\mat{L},\mat{\Lambda})$ possible pairs, which leads to optimal performance  in our scenario. For the sake of clarity, we show the results of our joint optimization over $\mat{L}$ and $\mat{\Lambda}$ as functions of one parameter at time. 

  Fig.  \ref{fig:L1varying_N200_Safearea} depicts the mean utility function as a function of  the number of burst slots  in the first class  $L_1$, for  $\Lambda_1(x)=\Lambda_2(x)= \Lambda^c(x)$, $N=200$ transmission slots, and source importance $(w_1,w_2)=(0.7,0.3)$. For each value of $L_1$, the number of messages in the second class $L_2$ is optimized with the different algorithms. Analogously, in Fig. \ref{fig:Different_Lambda_N200_w07}, the optimal utility function is depicted for different  polynomial distributions   $\Lambda_1(x)$, where the indexes on the x-axis follow the order in Table  \ref{tab:Lambda}. For each distribution $\Lambda_1(x)$,  $\Lambda_2(x)$ and $(L_1,L_2)$ are optimized with the algorithms under comparison. For both figures, the   optimization based on the theoretical ON region does not match   the optimal results achieved by exhaustive search. This is due to the mismatch between theoretical and actual ON region, as described above. However, we can observe that   the   optimization  based on the safe ON region in Algorithm 2 achieves a performance that approaches the optimal one. 
 We also observe that by increasing the number of  BNs dedicated to the most important class (i.e., $L_1$) the global utility function increases (Fig. \ref{fig:L1varying_N200_Safearea}). This is expected since a more prioritized transmission strategy is offered for large values of $L_1$. 
 If we rather look at the evolution of the global utility function as a function of several transmission probability distributions in  Fig. \ref{fig:Different_Lambda_N200_w07}, we notice an almost constant behavior. This is justified by the fact that  the MAC frame can be efficiently utilized and by properly tuning $\Lambda_2(x), L_1$, and $L_2$ for each value of $\Lambda_1(x)$, the system    achieves a large enough   throughout  to be in the floor region of the logarithmic utility function. 
   
\begin{figure}[t]
\begin{center}
\subfigure[$2$ source classes, $\Lambda_1(x)=\Lambda^e(x)$, and $\Lambda_2(x)= \Lambda^a(x)$.]{
\includegraphics[width=0.44\linewidth,  draft=false]{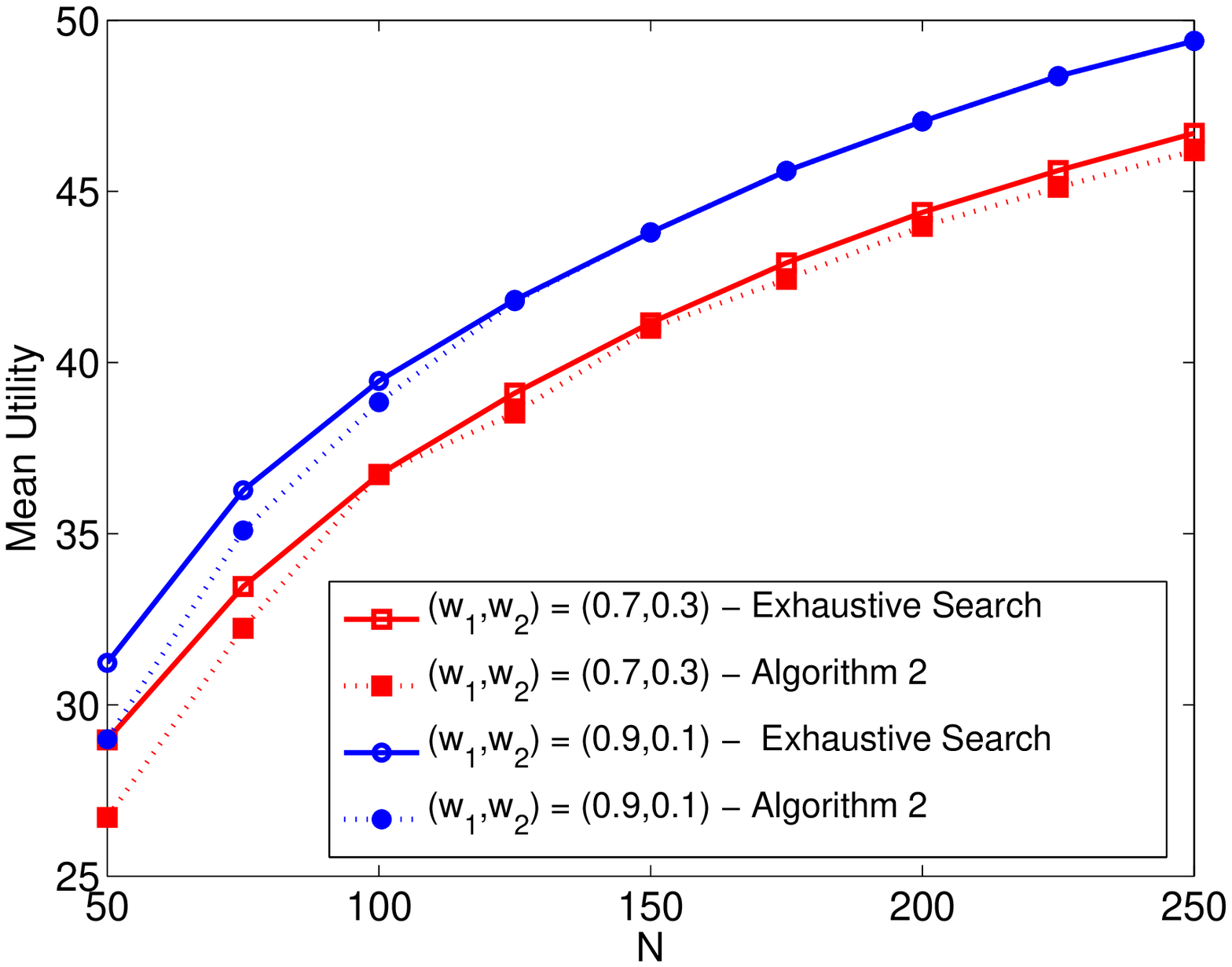} 
\label{fig:Utility_OPT_vs_N}
}
\subfigure[$4$ source classes, $\Lambda_k(x)=\Lambda^e(x), \forall k$.]{
\includegraphics[width=0.44\linewidth,  draft=false]{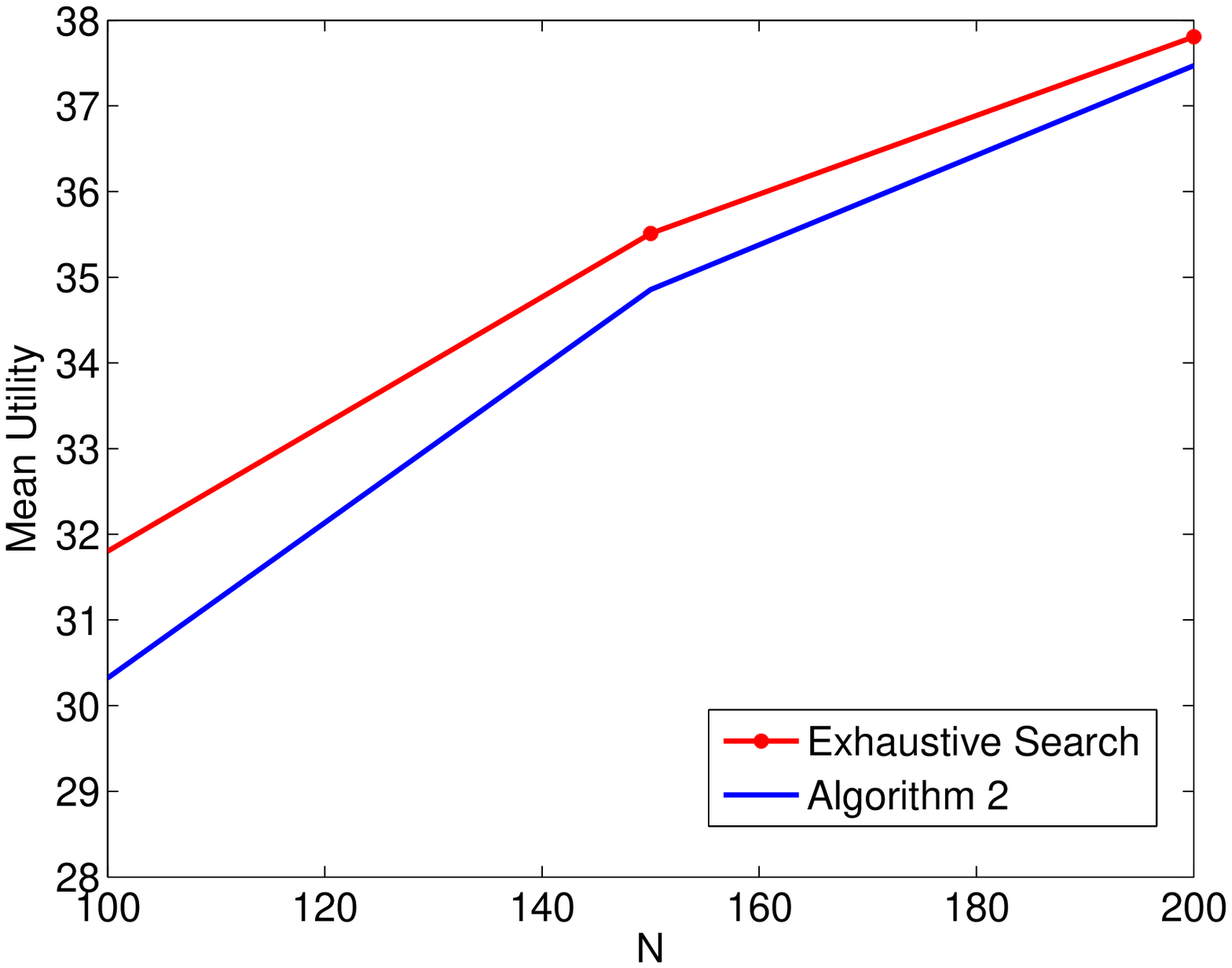} 
\label{fig:meanUtility_OPT_4classes}
}
\caption{Mean utility vs. the MAC frame size $N$, when $\{L_k\}$ is optimized in the case of $2$ and $4$ source classes.  }
\end{center}
\end{figure}


We further evaluate the  performance of the optimization method proposed in Algorithm 2 in different system settings and   provide in Fig.  \ref{fig:Utility_OPT_vs_N}   the mean utility function  as a function of the number of transmission slots $N$, given that two classes are considered with $\Lambda_1(x)= \Lambda^e(x)$, and $\Lambda_2(x)=  \Lambda^a(x)$. The optimization consists in finding the best pair $(L_1,L_2)$ for each value of $N$,  for two different pairs of weight $(w_1,w_2)$. The optimal performance obtained via exhaustive search is  compared to    the one  obtained with  Algorithm 2.  In all cases, we see that the Algorithm 2 reaches  mean utility scores that   are almost optimal.   As expected, we also observe that the global utility function increases with $N$. 
Finally, we run experiments in a larger system with  $K=4$ classes with importance   $(w_1,w_2,w_3,w_4) = (0.6, 0.2, 0.1, 0.1)$. The overall utility function is again provided as a function of the MAC frame size  $N$ when the number of messages is optimized with Algorithm 2 and Exhaustive search. The polynomial distribution is set to $\Lambda_k(x)= \Lambda^e(x)$ for all $k=[1,4]$. The results in Fig. \ref{fig:meanUtility_OPT_4classes} confirm the good match of the performance of our heuristic-based optimization algorithm with the optimal performance. 
We  notice again that  the achieved overall utility function increases with the MAC frame size $N$, as expected. 

 In conclusion, we have shown that the SIC theory can be applied to practical optimization problems, namely resource allocation strategies for prioritized sources.  An optimal resource allocation should be evaluated  either by finite-length analysis (which is unfortunately    not available in the literature for IRSA cases) or by simulation results (not feasible in terms of computational complexity). Thus, we have proposed an effective heuristic solution that is   practical to use and  achieves performance approaching the optimal one.

\section{Conclusions}
\label{sec:conclusions}
We have proposed prioritized new IRSA transmission strategies for systems with sources with different levels of importance. We have derived   a theoretical study of the  system performance in IRSA schemes with  heterogenous sources and analyzed   the asymptotic message error probability per class,   as well as the   global stability conditions. We have then proposed  a new  optimization problem aimed at finding the best transmission strategy in prioritized IRSA, in terms of both the replication probability and the  source rate per class. A carefully designed heuristic-based algorithm has also been developed in order to optimize the transmission strategy in realistic conditions.  Simulation results have validated our theoretical analysis and    demonstrated the gain of the proposed prioritized strategy.  
The proposed solution is practical and yet accurate,   achieving performance close to the optimal one.  This work provides the  main theoretical and practical tools for a system designer to optimally select transmission strategies for prioritized sources communicating to a common base station in an uncoordinated way.  

\bibliographystyle{IEEEtran}
\bibliography{LDPC_SlottedAloha}

\end{document}